\documentclass[a4paper,12pt,nohyper]{JHEP3}
\usepackage{epsfig}
\usepackage{amsmath,latexsym,amssymb}
\usepackage{graphicx}
\usepackage[latin1]{inputenc}
\usepackage{braket}

\usepackage{framed}

\def\be{\begin{equation}}
\def\ee{\end{equation}}
\def\ba{\begin{eqnarray}}
\def\ea{\end{eqnarray}}

\newcommand{\apr}{\alpha'}
\newcommand{\hapr}{{\apr \over 2}}

\newcommand{\bpartial}{ \bar{\partial}}
\newcommand{\bz}{\bar{z}}
\newcommand{\bw}{\bar{w}}

\newcommand{\tg}{\tilde{g}}
\newcommand{\tG}{\tilde{G}}
\newcommand{\tB}{\tilde{B}}
\newcommand{\tX}{\tilde{X}}
\newcommand{\hp}{\hat{p}}
\newcommand{\hk}{\hat{k}}
\newcommand{\he}{\hat{\epsilon}}
\newcommand{\bmu}{\bar{\mu}}
\newcommand{\bnu}{\bar{\nu}}
\newcommand{\pol}{\epsilon}
\newcommand{\hs}{\hat{s}}
\newcommand{\hht}{\hat{t}}
\newcommand{\hz}{\hat{z}}
\newcommand{\halft}{{t\over 2}}
\newcommand{\kn}{\kappa_N}

\def\a{\alpha}

\def\b{\beta}

\def\G{\Gamma}

\def\e{\epsilon}

\def\P{\Pi}

\title{\boldmath Pomerons and BCFW recursion relations for strings on D-branes \unboldmath}

\author{Angelos~Fotopoulos${}^{\spadesuit}$ and Nikolaos~Prezas${}^{\diamondsuit}$\\

  \begin{itemize}
    
  \item  Dipartimento di Fisica Teorica
    dell Universita di Torino and\\
  INFN Sezione di Torino,\\
  via P. Giuria 1, I-10125 Torino, Italy.
  
  \item Albert Einstein Center for Fundamental Physics, \\
        Institute for Theoretical Physics, University of Bern, \\
        Sidlerstrasse 5, CH-3012 Bern, Switzerland.
  \end{itemize}

\bigskip

E-mail: \email{foto@to.infn.it, prezas@itp.unibe.ch}}

\abstract{We derive pomeron vertex operators for bosonic strings
and superstrings in the presence of  D-branes. We demonstrate how they can be
used in order to compute the Regge behavior of string amplitudes
on D-branes and the amplitude of ultrarelativistic D-brane
scattering. After a lightning review of the BCFW method, we
proceed in a classification of the various BCFW shifts possible in
a field/string theory in the presence of defects/D-branes. The
BCFW shifts present several novel features, such as the
possibility of performing single particle momentum shifts, due to
the breaking of momentum conservation in the directions normal to
the defect. Using  the pomeron vertices  we show that superstring
amplitudes on the disc involving both open and closed strings
should obey BCFW recursion relations. As a particular example, we
analyze explicitly the case  of $1\to 1$ scattering of level one
closed string states off a D-brane. Finally, we investigate
whether the  eikonal Regge regime conjecture holds in the presence
of D-branes.}

\preprint{DFTT 14/2010}

\begin{document}

\setcounter{figure}{0} \setcounter{table}{0}

\setcounter{footnote}{0}
\renewcommand{\thefootnote}{\arabic{footnote}}
\setcounter{section}{0}

\section{Introduction and summary of results} \label{intro}

Recently, there has been remarkable progress in exploring the
properties of the S-matrix for tree level scattering amplitudes in gauge and gravity theories.
Motivated by Witten's twistor formulation of ${\cal N}=4$ Super--Yang--Mills (SYM)
\cite{Witten:2003nn}, several new methods have emerged which allow
one to compute tree level amplitudes.
The Cachazo--Svrcek--Witten (CSW)  method  \cite{Cachazo:2004kj} has
demonstrated how one can use the maximum helicity violating (MHV)
amplitudes of \cite{Parke:1986gb} as field theory vertices to
construct arbitrary gluonic amplitudes.

Analyticity of gauge
theory tree level amplitudes has lead to the Britto--Cachazo--Feng--Witten (BCFW) recursion relations
\cite{Britto:2004ap, Britto:2005fq}.
Specifically, analytic continuation of external momenta in a
scattering amplitude allows one, under certain assumptions, to
determine the amplitude through its residues on the
complex plane. Locality and unitarity require that the residue at
the poles is a product of lower-point  amplitudes.
Actually the  CSW construction turns out to be a particular application
of the BCFW method \cite{Risager:2005vk}.

The power of these new
methods extends beyond computing
 tree level amplitudes. The
original recursion relations for gluons \cite{Britto:2004ap}
were inspired by the
infrared (IR) singular behavior of ${\cal N} = 4$ SYM. Tree
amplitudes for the emission of a soft gluon from a given
n-particle process are IR divergent and this divergence is
cancelled by IR divergences from soft gluons in the 1-loop
correction. For maximally supersymmetric theories these IR
divergences suffice to determine fully the form of the 1-loop
amplitude. Therefore, there is a direct link between tree level
and loop amplitudes. Recently there has been intense investigation
towards a conjecture \cite{ArkaniHamed:2009dn, ArkaniHamed:2009si}
which relates IR divergences of multiloop amplitudes with those
of lower loops, allowing therefore the analysis of the full
perturbative expansion of these gauge theories\footnote{Very
recently there has been a proposal for the amplitude integrand at
any loop order in ${\cal N}=4$ SYM \cite{ArkaniHamed:2010kv} and a
similar discussion on the behavior of loop amplitudes under BCFW
deformations \cite{Boels:2010nw}.}. These new methods have
revealed a deep structure hidden in maximally supersymmetric gauge
theories \cite{ArkaniHamed:2008gz} and possibly in more general
gauge theories and gravity.

The recursion relations of \cite{Britto:2005fq} are in the heart
of many of the aforementioned developments. Nevertheless it
crucially relies on the asymptotic behavior of amplitudes under
complex deformation of some external momenta. When the complex
parameter, which parametrizes the deformation, is taken to
infinity an amplitude should fall sufficiently fast so that there is no
pole at infinity\footnote{There has been though some recent
progress  \cite{Feng:2009ei,Feng:2010ku}
 in generalizing the  BCFW relations for theories with boundary
contributions.}. Although naive
power counting of individual Feynman diagrams seems to lead to
badly divergent amplitudes for large complex momenta, it is
intricate cancellations among them which result in a much softer
behavior than expected. Gauge invariance and supersymmetry in some
cases lies into the heart of these cancellations.

A very powerful
criterion has been developed in \cite{ArkaniHamed:2008yf}
and studied further in \cite{Cheung:2008dn}, which
allows one to infer, purely from the symmetries of the tree level
Lagrangian, which theories allow BCFW relations and for which
helicity configurations. Considering the deformed particles as
``hard"  which propagate in the ``soft" background created by
the undeformed particles, we can study the asymptotic behavior of
scattering amplitudes under large complex deformations. This
behavior is ultimately related to the ultraviolet (UV) properties of the theory
under consideration. For example, ${\cal N}=4$ SYM is a theory
with excellent UV behavior and the BCFW relations take very simple
form allowing one to determine fully all tree level amplitudes of
the theory for arbitrary helicity configurations \cite{Drummond:2008cr}.

It is natural to wonder whether these field theoretic methods can be applied
and shed some light into the structure of string theory amplitudes.
This is motivated, in particular, by the fact that the theory that plays
 a central role in the developments we described above, that is
${\cal N}=4$ SYM,  appears as the low-energy limit
of  string theory in the presence of  D3-branes.
In order to even
consider applying the aforementioned methods to string scattering
amplitudes, one needs as a first step to study their behavior for large complex
momenta. Since generally string
amplitudes are known to have excellent large momentum behavior, one
expects that recursion relations should be applicable here as
well. Nevertheless, one should keep in mind that although the
asymptotic amplitude behavior might be better than any local field
theory, the actual recursion relations will be quite more involved.
The reason is that 
they will require knowledge of an infinite set of on-shell string
amplitudes, at least the three point functions, between arbitrary
Regge trajectory states of string theory.

The study of the asymptotic behavior of string amplitudes under
complex momentum deformations was initiated in \cite{Boels:2008fc}
and  elaborated further in \cite{Cheung:2010vn,Boels:2010bv}. 
These works  established, using  direct study of the
amplitudes in parallel with  pomeron techniques, that both open and
closed bosonic and supersymmetric string theories have good
behavior asymptotically, therefore  allowing one to use the BCFW
method\footnote{Recently, pomerons have also been used in
\cite{Fotopoulos:2010ay} to advocate that BCFW relations exist in
higher spin theories
constructed as the tensionless limit of string theories.}. Moreover, it was
observed in \cite{Cheung:2010vn} that for the supersymmetric
theories the leading and subleading asymptotic behavior of open
and closed string amplitudes is the same as the asymptotic
behavior of their field theory limits, i.e.~gauge and gravity theories
respectively. This led to the eikonal Regge (ER) regime conjecture
\cite{Cheung:2010vn} which
 states that string theory amplitudes, in a region where some
of the kinematic variables are much greater than the string scale
and the rest much smaller, are
reproduced by their corresponding field theory limits. For bosonic
theories there is some discrepancy in some subleading terms
\cite{Boels:2010bv} which most probably can be attributed to the
fact that an effective field theory  for bosonic strings is
plagued by ambiguities due to the presence of tachyonic modes.

The purpose of the current work is to study string amplitudes
which involve both open and closed strings 
in the presence of D-branes. We will use the method of
pomeron operators since we believe it is  the most direct one
and it will allow us, as a byproduct, to derive a few interesting
results which apply beyond BCFW. In addition, the analysis
of BCFW deformations is general
and applies  to all situations
where a defect (brane) in spacetime interacts with
bulk modes.

In section \ref{CFT} we give a short review of the relevant CFT
machinery for string operators in the presence of a boundary on
the world-sheet. In section \ref{PomTech} we use the operator product expansion (OPE)
technology at hand to derive the pomeron operators, along the lines
of \cite{Brower:2006ea}, for D-branes in a flat background. The main
idea is to divide the string operators into two sets highly
boosted relatively to each other. The pomeron operator is
exchanged between the two sets of operators and exhibits the
typical Regge behavior expected for such a process. We study the
Regge behavior of the prototype amplitude for the scattering of
two closed string tachyons off a D-brane.

There are two
qualitatively different factorization channels of the string
amplitude on the world-sheet which lead to two different pomeron
operators. The first factorization, into a sphere and a disc, leads
to the closed string pomeron operator while  the second,
into two discs, leads to the open pomeron operator. This
factorization can appear even if one of the two sets of operators
has a single closed string operator. This is unlike the usual
purely closed string theory analysis which requires at least two
operators for each set. This is a novel feature due to the
presence of the D-brane.

Subsequently,   section 2 contains two applications.
First, we compute a mixed open-closed string amplitude on
the disc applying the pomeron machinery. This example demonstrates
an important point. The  field theory limit factorization of string 
amplitudes cannot always be reproduced by a corresponding 
Regge channel factorization.
This is unlike the pure open string theory where 
Regge channels due to two highly-boosted gluons
coincide with the two particle factorization channels of
gauge theory amplitudes. Actually for some amplitudes there is no
Regge type behavior at all for the world-sheet factorization which
leads to the massless poles. This will have profound consequences
for the ER regime conjecture in the presence of D-branes.

The
second application deals with ultrarelativistic scattering of
D-branes. Unlike the previous examples, where pomerons were used
for computing scattering of highly boosted string states off  a
D-brane, in this case we use the pomeron technique to compute the
scattering of one D-brane off another. This way we can reproduce
the result of \cite{Bachas:1995kx}. We conclude this section by
giving the explicit formulas for the pomeron operators for level
one states of both bosonic and superstring theories. There are
three different pomerons for each case. The open string pomeron
from the OPE  of two open string states on the disc, the closed
string pomeron from the OPE of two closed strings on the sphere
and a tadpole pomeron operator from the OPE of a closed string
with its image as it approaches the boundary of the disc
world-sheet.

The results of this section can be useful for holographic
computations as well. One would need to extend the pomeron
operators in a curved, asymptotically AdS background along the
lines of \cite{Brower:2006ea}. Open strings attached to D-branes
model quarkonium states and closed strings glueballs.. Therefore
pomeron operators for open-closed string amplitudes would be
useful in studying quarkonium-glueball scattering in the Regge
regime. Furthermore, D-branes correspond holographically to
non-perturbative states of the boundary gauge theory. For D-brane
scattering we would need the corresponding boundary states for
highly boosted D-branes moving in the curved background. We leave
these interesting applications for future investigations.

In section \ref{BCFWrev} we give a lightning review of the BCFW
method for field theories. We continue with an analysis of the
possible BCFW shifts for scattering amplitudes in the presence of
defects. The analysis is general and does not rely on string
theory. It can be used for field theories in the presence of
defects such as  brane-world theories. There is a plethora of
multi-particle BCFW shifts. We concentrate on the two simplest
cases the two particle and single particle shifts. The case of the
two particle shift is momentum conserving in all
 directions and encompasses the standard BCFW shift for both bulk and brane modes. The single particle shift, special to
theories where a defect is present, includes shifts which violate
momentum conservation in the directions transverse to the defect
and applies only to bulk modes. The single particle shift is the
one which leads to the tadpole pomeron operators of the previous
section.

In section \ref{Pom} we use the results of sections \ref{PomTech}
and \ref{BCFWrev} to derive the behavior of disc scattering amplitudes
under two particle momentum shifts for open and closed strings
and the one particle momentum closed string shift. The pattern for the two
particle shifts is the same as that for gauge theory and gravity
respectively. The one particle shift leads to a behavior similar
to the two particle open string shift as dictated by the Kawai--Lewellen--Tye (KLT)
\cite{Kawai:1985xq} relations. For superstring amplitudes the
behavior derived is the same as the field theory analysis for
gauge and gravity theories. We conclude the section with an
explicit example where we compute the BCFW  behavior for the
superstring scattering amplitude on the disc of two level one
closed string states for both the two particle and one particle
shifts. We verify that indeed our pomeron analysis gives the
correct behavior.

Section \ref{FT} makes an attempt to identify a field theory whose
behavior under BCFW shifts is the same as that of the string
theory amplitudes, therefore extending the ER regime conjecture to
the case of D-branes. We do not succeed in identifying  such a
field theory and there are several good reasons why the  conjecture
might not be applicable for string amplitudes involving D-branes.
The first observation is that tree level field theory diagrams
describing  exchange of a bulk field between D-branes correspond
to loop amplitudes in the string theory side. For instance, a
closed string exchanged between D-branes is related through
world-sheet duality to the annulus (one-loop) amplitude of open
strings. 

Another crucial point, as mentioned before, is that Regge
channel factorization does not always agree with the field theory
limit factorization of the amplitude.  In addition, there are
amplitudes whose behavior under certain BCFW shifts is not of
Regge type. Nevertheless, they
have the behavior predicted by the Dirac--Born--Infeld (DBI)
effective field theory.  However, the field
theory diagrams constructed using the DBI vertices corresponding to 
non-Regge amplitudes, lead into disagreement with
the pomeron analysis for amplitudes with Regge behavior.

 These problems might have been expected since, as pointed out in
\cite{ArkaniHamed:2008yf}, in a gauge theory with higher
dimension operators such as  $(F_{\mu \nu})^n, \ n>2$ the nice BCFW behavior of Yang--Mills (YM)
theory is spoiled. This dictates that we should take the low-energy limit of
the DBI action which eliminates all the aforementioned higher
dimension operators and leaves only the SYM action.
Moreover,
when we consider amplitudes with closed and open strings on
D-branes with Regge type behavior under BCFW shifts, we
find out that they cannot be reproduced by keeping closed string
couplings in the DBI action. Since these couplings are
proportional to Newton's constant $\kn$, they are higher dimension
operators and in the spirit of the discussion above should be
eliminated. We conclude that the ER regime conjecture might work only
in the decoupling limit of the D-brane theory. It would be
interesting to investigate if this conclusion has any implications for
high energy scattering in holographic backgrounds.

\section{Conformal field theory with D-branes}\label{CFT}

In this paper we will consider the bosonic string  and the NS sector
of the superstring.  In this section we summarize for convenience the relevant notions  from CFT that
we will employ  (see, for instance, \cite{Garousi:1996ad, Hashimoto:1996bf, Fotopoulos:2001pt}).

In spacetime Dp-branes are represented as
 static (p+1)-dimensional defects. As a result of these defects we must impose different
 boundary conditions on the world-sheet boundary to coordinates tangent and
 normal to the D-brane
  \begin{eqnarray} \label{NeuDir}
  \partial_{\perp}X^{\alpha}\mid_{\partial \Sigma}=0\ , \nonumber \\
  X^{i}\mid_{\partial \Sigma}=0\ .
  \end{eqnarray}
  The lower case Greek indices  $ \alpha=0,1 \dots ,p$
  correspond to directions parallel to the D-brane while
  the lower Latin ones $ i=p+1,
  \dots, 9$ to normal coordinates. The boundary conditions (\ref{NeuDir}) are respectively Neumann
and Dirichlet.

The string vertex  operators of a closed string are factorized in holomorphic and antiholomorphic parts
and they take the following schematic form
\begin{eqnarray}\label{GenOper}
V_{(s, \bar s)}(z, \bar{z})\sim \epsilon :V_{(s)}(z): \,
:V_{(\bar s)}(\bar{z}):
\end{eqnarray}
where $\epsilon$ is a polarization tensor.
For the superstring $ s$
denotes the superghost charge or equivalently the picture in which
the operator is in. The total superghost charge on the disk is
required to be $Q_{sg}=-2$ as a consequence of
superdiffeomorphism invariance.

The holomorphic parts of the vertex operators for bosonic string tachyons and level 1 states
carrying momentum
$p$ are
\begin{eqnarray}\label{bosoper}
V^{T}(z;p)&=&  e^{ i p \cdot
X(z)}\ , \nonumber \\
V^{\mu}(z;p)&=&  \sqrt{{2\over \apr}}
\partial X^{\mu}(z) e^{ i p \cdot X(z)}
\end{eqnarray}
respectively. For example, the vertex operator for a closed string tachyon reads
 $V^{T_c}(z, \bz;p)= \ V^{T}(z;p)  \bar{V}^{T}(\bz;p)$. Notice that the vertex operators for
 open string tachyons and gluons take a form similar to the above expressions.

The holomorphic parts of the vertex operators corresponding to massless states (level 1) of the
 closed superstring are given by
\begin{eqnarray}\label{PicOper}
V_{(-1)}^{\mu}(z;p)&=& e^{-\varphi(z)} \psi^{\mu}(z) e^{i p \cdot
X(z)}\nonumber \\
V_{(0)}^{\mu}(z;p)&=& \sqrt{{2\over \apr}}\Big( \partial X^{\mu}(z) + i  \hapr  p
\cdot \psi(z) \psi^{\mu}(z)\Big)e^{ i p \cdot X(z)}
\end{eqnarray}
in the -1 and 0 picture respectively with
similar expressions for the anti-holomorphic part. As usual $\varphi$ is the bosonized superconformal
ghost.

The expectation values of string vertices are found using the
following correlators:
\begin{eqnarray}\label{HolCor}
\left \langle X^{\mu}(z) X^{\nu}(w) \right \rangle&=& -\hapr \ \eta^{\mu \nu}\log (z-w) \ ,\nonumber \\
\left\langle \psi^{\mu}(z) \psi^{\nu}(w) \right\rangle &=& - \frac{\eta^{\mu
\nu}}{z-w}\ , \\
\left\langle \varphi(z) \varphi(w) \right \rangle &=& - \log (z-w)\ . \nonumber
\end{eqnarray}
The Minkowski metric is $\eta^{\mu\nu}={\rm diag}\{-1,+1,\ldots,+1\}$.
Because of the boundary conditions we have non-trivial correlators
between the holomorphic and the antiholomorphic parts
\begin{eqnarray}\label{HolAntiCor}
\left \langle X^{\mu}(z) \bar{X}^{\nu}(\bar{w}) \right\rangle&=& - \hapr \
D^{\mu \nu}\log (z-\bar{w})\ ,
\nonumber \\
\left \langle \psi^{\mu}(z) \bar{\psi}^{\nu}(\bar{w}) \right \rangle &=& -
\frac{D^{\mu \nu}}{z-\bar{w}}\ ,  \\
\left \langle \varphi(z) \bar{\varphi}(\bar{w}) \right \rangle &=& - \log (z-\bar{w})
\nonumber
\end{eqnarray}
where $ D_{\nu}^{\mu}$ is a diagonal matrix with +1 for directions
tangent to the world-volume and -1 for the normal directions.
\begin{equation}\label{corrD1}
\partial X^\mu (z) :e^{i p \cdot X(w)}: \; \sim -i   p^\mu \frac{\alpha'}{2} \frac{: e^{i p \cdot X(w)}:}{z-w}\ ,
\end{equation}
\begin{equation}
:e^{i p_1 \cdot X(z)} : \; :e^{i p_2 \cdot X(w)} : \; \sim    (
z-w)^{\frac{\alpha'}{2} p_1\cdot p_2} :e^{i p_1 \cdot X(z)} e^{i
p_2\cdot X(w)}: \ .
\end{equation}

 In some cases it will be more convenient to use the
correlators on the disc \cite{klebanov} rather than on the
half-plane:
\begin{eqnarray} \label{corrD2}
\left \langle X^{\mu}(z) \bar{X}^{\nu}(\bw) \right \rangle&=& -\hapr\ D^{\mu
\nu} \log (1-z \bw)\ , \nonumber \\
 \left \langle \psi^{\mu}(z) \bar{\psi}^{\nu}(\bw) \right \rangle &=&
\displaystyle  i \frac{D^{\mu \nu}}{1-z \bw} \ .
\end{eqnarray}

At this point it is useful to define two projection matrices
$ V_{\nu}^{\mu}$ and $ N_{\nu}^{\mu} $ which project on the
tangent and normal directions of the brane respectively and satisfy
\begin{eqnarray}\label{VN}
D_{\mu \nu}= V_{\mu \nu} - N_{\mu \nu}\ , \quad
\eta_{\mu \nu} = V_{\mu \nu} + N_{\mu \nu} \ .
\end{eqnarray}
It is important to notice that momentum is conserved only
along the tangent directions. Therefore, if we have a set of strings scattering with
momenta $p_I^\mu$, momentum conservation takes the following form:
\begin{equation}\label{momcons}
\sum_{i=1}^n (p_I+ Dp_I)^\mu =0\ .
\end{equation}
 Extending the definition of the fields to the whole complex plane
 \cite{Hashimoto:1996bf} we can write all our vertices in terms of left moving string
operators by making the substitutions
\begin{eqnarray} \label{substitute}
\bar{X}^{\mu}(\bar{z}) \rightarrow D^{\mu}_{\nu} X^{\nu}(\bar{z})\ ,
\quad \bar{\psi}^{\mu}(\bar{z}) \rightarrow D^{\mu}_{\nu}
\psi^{\nu}(\bar{z})\ , \quad
\bar{\varphi}(\bar{z}) \rightarrow
\varphi(\bar{z}) \nonumber
\end{eqnarray}
where $ z \in \cal{H}^{+}$. This way we use the standard correlators
(\ref{HolCor}) in string amplitude computations.
 After these replacements we write for the case of a graviton with polarization $\epsilon_{\mu \nu}$
 in the superstring
\begin{eqnarray} \label{GravOper}
V_{(s,\bar s)}(z, \bar{z};p)= (\epsilon D)_{\mu \nu} :V_{s}^{\mu}(z;p): \,
:V_{\bar s}^{\nu}(\bar{z};Dp):
\end{eqnarray}
Similar manipulations allow us to write the open string vertex
operators for Neumann and Dirichlet conditions in the -1 and 0 picture as
\cite{Hashimoto:1996bf}
\begin{eqnarray} \label{OpenOper}
V_{(-1)}^{\mu}(x;2k)&=& e^{-\varphi(x)} \psi^{\mu}(x) e^{ i 2k \cdot
X(x)}\ , \nonumber \\
V_{(0)}^{\mu}(x;2k)&=&\sqrt{{2\over \apr}}\Big( \partial X^{\mu}(x) +
i \apr k \cdot \psi(x) \psi^{\mu}(x) \Big)e^{ i 2k \cdot
X(x)}
\end{eqnarray}
with $x \in \mathbb{R}$
and the momentum $k^{\mu} $ restricted
to be tangent to the brane. The vertex operators (\ref{OpenOper})
have momenta $2k$ in the exponent and a slightly  different form in the
zero picture as compared to (\ref{PicOper}).
This is due to the fact
we use the correlators (\ref{HolCor}) and (\ref{HolAntiCor}) or  (\ref{corrD2})
which are for closed strings and do not take care of the double
subtraction needed for normal ordering operators on the boundary
of the disc \cite{Hashimoto:1996bf}\footnote{ Usually in the
literature one sets for simplicity $\alpha'=2$ for both closed
and open strings. We keep here $\apr$ explicitly since it is
useful in order to track leading momentum contributions in our
formulas. }. This will be important later when comparing our
pomeron operators with those which have appeared recently in the
literature.

We have now all the ingredients to compute an amplitude between $n$ open strings
carrying momenta $k_I, I=1,\ldots,n$ and $m$ closed strings with momenta $p_J, J=1,\ldots,m$.

\begin{eqnarray} \label{StrAmp}
{\cal A}\left(\{k_I\}, \{p_J\}\right) =  \displaystyle\int_{D_2} \frac{ d^2 \!z_{J} d  x_{I}}{V_{CKG}}
 \left \langle \prod_{I=1}^{n} : V(x_{I};k_I): \; \prod_{J=1}^{m} :V(z_{J},
\bar{z}_{J};p_J): \, \right \rangle
\end{eqnarray}
with $V_{CKG}$ being the volume of the conformal Killing group. The open string momenta
can only be in directions tangent to the brane and momentum conservation reads\footnote{The projector
$V$ in the formula below should not be confused with a vertex operator!}
\begin{equation}
\sum_{I=1}^n k_I + \sum_{J=1}^m V\cdot p_J =0\ .
\end{equation}

\section{Pomeron techniques in the presence of D-branes}\label{PomTech}

Pomeron vertex operator for string amplitudes first appeared
in \cite{Brower:2006ea}. We will use their method to derive
pomeron vertex operators in the presence of boundary states.
We will consider the scattering of two  closed string tachyons off a
D-brane and  we will demonstrate the  Regge behavior of
this amplitude. Subsequently, we will use the OPE of tachyon operators
in bosonic string theory and we will compute the pomeron vertex
that reproduces the aforementioned Regge behavior. The purpose of
computing the Regge limit behavior for this prototype amplitude is
that it will enables us to identify two qualitatively different
pomeron channels due to the presence of the boundary on the
world-sheet.

\subsection{Pomeron operators for tachyons}\label{PomTac}
We consider the amplitude on the disc $D_2$ of  two closed string tachyons with
momenta $p_1$ and $p_2$. Using the correlators
(\ref{corrD2}) and after dividing by $V_{CKG}$ we obtain
\begin{eqnarray}\label{2TD}
{\cal A}(p_1, p_2) &\sim& \int_{D_2}  \; d^2 w \; \langle :V^{T_c}(w,
\bw;p_1): :V^{T_c}(0, 0;p_2): \, \rangle_{D_2} \nonumber \\
&\sim& \int_{D_2} d^2 w (1- |w|^2)^{\apr s / 2} |w|^{-\apr t / 2}
\end{eqnarray}
where the kinematic variables are  $s= p_1 \cdot D \cdot p_1=p_2 \cdot D \cdot p_2$ and $ t=-2 p_1 \cdot p_2$.
As is usual in the computation of string amplitudes, we assume that the kinematic
variables take appropriate values where the integrals are convergent and
once we are done with the integrations we can analytically continue to the
physical region. For the integral in (\ref{2TD}) convergence requires that $s>0$ and $ t<0$.

Following the saddle point method of  \cite{Brower:2006ea}
we are able to identify
the region of the world-sheet where the two different channels
dominate:
\begin{eqnarray}\label{saddle}
&& {\rm t-channel} \ : \ -t << s \longrightarrow |w|^2 \sim  -\frac{t}{s} \ ,  \nonumber \\
&& {\rm s-channel} \ : \ s << -t \longrightarrow |w|^2 \sim 1 \ .
\end{eqnarray}
The two channels are named after the subleading momentum invariant and
correspond to distinct pomeron states giving the leading
contribution to scattering where a set of operators is highly
boosted with respect to the rest of the operators in the path
integral. For scattering on the disc they correspond to two distinct factorization
channels on the world-sheet: closed string (t-channel) and open string
(s-channel) (see, for instance, \cite{Fotopoulos:2002wy}).
 In the ensuing OPE analysis we will use
hatted symbols to indicate the momenta that are highly boosted.
This will result in some kinematic invariants being larger than others,
as for example in (\ref{saddle}), and we will use hatted symbols for them as well.

More specifically, for amplitudes on the sphere or purely open string
amplitudes on the disc, kinematic invariants built from products
of boosted or unboosted momenta are unhatted (small) while those
from cross-products of highly boosted momenta with unboosted ones
are hatted (large). We refer the reader to \cite{Brower:2006ea}
for further details. A subtlety is that in the presence of D-branes
it is possible that kinematic invariants built
only from boosted momenta are large. We shall see shortly such
examples. In the present pomeron discussion this notation is not
particularly useful but we will keep it in order to be in accord  with the
BCFW analysis which will follow later on.

\subsubsection{t-channel}

Let us first  study the integral (\ref{2TD}) in the t-channel
region. The result is
\begin{eqnarray}\label{t-int}
\int_{S^2} d^2 w \ e^{\apr \hat{s} / 2  \log(1-|w|^2)}
|w|^{-\apr t /2 } &\sim& \int_{S^2} d^2 w \ e^{-\apr \hat{s} /2 \
|w|^2} |w|^{-\apr t/2} \nonumber \\
&\sim&   \Gamma \left(1- {\apr t \over 4}\right) \ \left({\apr
\hat{s}\over 2}\right)^{{\apr t \over 4}-1}\ .
\end{eqnarray}
Notice that we are integrating on a sphere since the region $|w|<<1$ is
far from the boundary of the disc.

The corresponding pomeron vertex operator  can be
extracted from the OPE
\begin{eqnarray}\label{tOPE}
:e^{i \hp_1 \cdot X(w, \bw)}: \; : e^{i \hp_2\cdot X(0, 0)}: &\sim&
|w|^{-\apr t / 2} : e^{i \hp_1\cdot
X(w) + i \hp_2\cdot X(0)}: : e^{i \hp_1 \cdot \bar X(\bw) + i\hp_2 \cdot \bar X(0)}:  \\
&\sim& |w|^{-\apr t /2 } : e^{i (\hp_1 +\hp_2)\cdot X(0) + i
w\hp_1\cdot
\partial X(0)}: : e^{i (\hp_1+\hp_2)\cdot \bar X(0) + i\bw \hp_1
\cdot\bpartial \bar X(0)}:  \ . \nonumber
\end{eqnarray}
As in \cite{Brower:2006ea}  we have kept the terms subleading in $w$ and $\bar w$.
Moreover,
we have normal ordered separately the  holomorphic $X$ and
antiholomorphic $\tX$ fields since when this operator approaches
the boundary there will be non-trivial correlators between them
due to (\ref{corrD2}). Integrating
over $w$
 we extract
the pomeron vertex\footnote{Notice that compared to the pomeron
vertex operator of (3.6) in \cite{Brower:2006ea} we have an extra
phase. This is in agreement with (3.20) and
(3.21) of the same paper.}
\begin{equation}\label{tPom1}
{\cal V}^{T_c T_c}= \P^c(\apr t) :e^{i p \cdot X(0)} (i
\hp_1 \cdot
\partial X)^{{\apr t \over 4}-1} : \; : e^{i   p \cdot \tX(0)} (i \hp_1 \cdot
\bpartial \bar X)^{{\apr t \over 4}-1} :
\end{equation}
where $p=\hp_1+\hp_2$ and the closed pomeron propagator reads
\begin{equation}\label{tPP}
\Pi^c(\apr t )= {\Gamma\left(1-{\apr t\over 4}\right)\over
\Gamma\left({\apr t \over 4}\right)} e^{i \pi \left(1-{\apr t\over
4}\right)}.
\end{equation}
 Actually, to be closer to the spirit of
\cite{Brower:2006ea} we should expand the OPE symmetrically in the positions of
the two operators, therefore obtaining
\begin{equation}\label{tPom2}
{\cal V}^{T_c T_c}= \Pi^c(\apr t) :e^{i p \cdot X(0)} (i Q
\cdot
\partial X)^{{\apr t \over 4}-1} : \; : e^{i p \cdot \bar X(0)} (  i Q \cdot
\bpartial \bar X)^{{\apr t \over 4}-1} :
\end{equation}
where $Q= { \hp_1-\hp_2 \over  2}$.
This  operator satisfies the physical state conditions of the
Virasoso algebra
\begin{eqnarray} \label{tPomphys}
 L_0 {\cal V}^{T_c T_c}&=& {\apr \over 4} p^2 + N-1= {\apr \over
4} (\hp_1+\hp_2)^2 + \left({\apr \over 4} t-1\right)-1=0\ , \nonumber \\
 L_1 {\cal V}^{T_c T_c}&\sim& Q \cdot p =0\ .
\end{eqnarray}

Taking the expectation value of this operator on the disc $D_2$
and using momentum conservation along the D-brane $(p+Dp)^\mu=0$,
yields
\begin{eqnarray}\label{2Tpomcl}
\left\langle {\cal V}^{T_c T_c}(0) \right\rangle_{D_2} &\sim& \Pi^c(\apr t) \
\left({\apr t\over 4}-1\right)! \ \left(-{\apr \over 2} Q\cdot D
\cdot Q\right)^{{\apr t\over 4}-1} \nonumber \\
&\sim&
 \Gamma\left(1-{\apr t\over
4}\right) \ \left({\apr \hat{s}\over 2}\right)^{ {\apr t\over 4}
-1}
\end{eqnarray}
which indeed agrees with (\ref{t-int}). We have used that $Q\cdot D\cdot Q =s -\frac{t}{4}+
\frac{m^2_{T_c}}{2}$, where $m^2_{T_c}=-4/\alpha'$ is the tachyon squared-mass, and so
 $Q\cdot D\cdot Q \sim s$ in the kinematic regime under consideration.

\subsubsection{s-channel}

Now let us proceed with the s-channel region. In this case (\ref{2TD}) becomes
\begin{eqnarray}\label{s-int}
\int_{D^2} d^2 w \ e^{-\apr \hht /4  \log|w|^2}
(1-|w|^2)^{\apr s /2} &\sim& \int_{D^2} d^2 w \ e^{-\apr \hht /4
(|w|^2-1)} (1-|w|^2)^{\apr s /2 }  \nonumber \\
&\sim&\Gamma \left({\apr
s\over 2}+1\right) \ \left(-{\apr \hht \over 4}\right)^{-{\apr s\over 2}-1}\ .
\end{eqnarray}
We have dropped the contribution of an incomplete gamma function
which  goes to zero for large $\alpha' t/4$.

Notice that this world-sheet region dominates when a single string
vertex operator approaches the disc boundary which we choose to be the
operator with momentum $\hp_1$. The OPE we need to consider this
time is between the holomorphic and anti-holomorphic pieces of the
corresponding vertex operator near the disc boundary
\begin{eqnarray}\label{sOPE}
:e^{i \hp_1\cdot X(w)}::e^{i \hp_1\cdot \bar X(\bw)}:&\sim&
(w-\bw)^{\apr s /2}:e^{i \hp_1
\cdot X(w)} e^{i D\cdot  \hp_1 \cdot X(\bw)}: \nonumber \\
 & \sim &(2iy)^{\apr s /2}:e^{i (\hp_1+D \cdot \hp_1)
\cdot X(x)+ i y \hp_1 \cdot(
\partial X- D\cdot \partial X)}:
\end{eqnarray}
where we used this time correlators on the upper half-plane and the
parametrization $w=x+iy$ with $y=0$ being the boundary. Performing
the integral over $y$ and using the identity $\hp+D\cdot \hp= 2V\cdot \hp$ results in
\begin{equation}\label{sPom1}
{\cal V}^{T_c}= \Pi^{o}(\apr s)\ (i  \hp_1 \cdot N \cdot
\partial X)^{-1-\apr s /2} \ e^{i(2 V\cdot \hp_1)\cdot X}\end{equation} or,
in the symmetric OPE expansion,
\begin{equation}\label{sPom2}
{\cal V}^{T_c}= \Pi^o(\apr s) \ (i  Q \cdot \partial
X)^{-1-\apr s /2} \ e^{i(2 k)\cdot X}\end{equation} where $2k=
\hp_1 + D\cdot \hp_1 =2 V \cdot \hp_1$ and $2Q=
\hp_1-D\cdot \hp_1 = 2 N\cdot \hp_1$ and the pomeron
propagator is
\begin{equation}\label{sPP}
\Pi^o(\apr s)=\Gamma\left({\apr s \over 2}+1\right)\ .
\end{equation}
The subscript  indicates that the pomeron vertex corresponds to
the factorization of the amplitude into a closed string tachyon  tadpole on
the D-brane connected via the pomeron with the rest of the
operators in the path integral. Such a behavior is possible due to
the non-conservation of momentum normal to
the D-brane and is one of the novelties of the present
analysis.

This operator is an open string pomeron vertex.
We can verify that the
physical state conditions are satisfied\footnote{The level $N$ in the first formula below
should not be confused with the projector $N$ in the second formula!}
\begin{eqnarray} \label{sPomphys}
 L_0 {\cal V}^{T_c}&=& \frac{\apr}{4} (2k)^2 + N-1=\apr  ( V\hp_1)^2 + \left(-1-{\apr \over 2} s\right)-1=0\ , \nonumber \\
 L_1 {\cal V}^{T_c}&\sim& Q\cdot k= N\cdot \hp_1 \cdot V \cdot \hp_1 =0\ .
\end{eqnarray}

To compare with the corresponding expression in (\ref{s-int}) it
is instructive to compute the amplitude in the symmetric manner
indicated in  \cite{Brower:2006ea}. This is the symmetric picture
of pomeron exchange which separates the boosted particles into two
sets and uses the pomeron vertex\footnote{Notice that in this
context the pomeron vertex (\ref{sPom2}) appears stripped of its
polarization vector $i Q^\mu$, much like in field theory the
intermediate gauge bosons polarizations are substituted with their
propagator.}  to connect the individual diagrams. We can indeed
compute the amplitude due to pomeron exchange between two
oppositely boosted closed string tachyons with momenta $\hp_1$ and
$\hp_2$
\begin{eqnarray}\label{2Tpomtad}
{\cal A}(\hp_1,\hp_2) &\sim& \left\langle V^{T_c}_1 \; {\cal
V}^{T_c}  \right\rangle_{D_2}\  \Pi^{o}(\apr s) \ \left\langle
V^{T_c}_2\; {\cal V}^{T_c}
\right\rangle_{D_2} \nonumber \\
 & \sim& \left( \sqrt \hapr N\cdot \hp_1\right)^{-1-\apr s /2} \cdot  \Pi^o(\apr s)   \ \left(
\sqrt \hapr N \cdot \hp_2\right)^{-1-\apr s/2}  \nonumber \\
&\sim&
\left(-{\apr \hht\over 4}\right)^{-1-\apr s /2} \Gamma\left(1+{\apr s\over 2}\right)
\end{eqnarray}
where we have suppressed the corresponding ghost contributions
which are needed to fix the individual CKG of the two discs. The
factors $\hapr$ in the computation of each disc amplitude appear
due to the normalization of the pomeron operators, as can be
derived by equation (3.12) of \cite{Brower:2006ea} for the open
pomeron and the contractions using (\ref{HolCor}) and
(\ref{HolAntiCor}) . Moreover we have used the fact that in this
regime we have $\hht=-2 \hp_1 \cdot \hp_2 \sim -2 \hp_1 \cdot
N\cdot \hp_2$, since $-2 \hp_1 \cdot V\cdot \hp_2 << \hp_1 \cdot N
\cdot \hp_2$ in the kinematic regime under consideration and the
expression in (\ref{2Tpomtad}) assumes contractions of the tensors
between the amplitude of the one disc and the other\footnote{A
more elegant way to put it is the $+/-$ plane formalism of
\cite{Brower:2006ea} but we will be a bit sloppy and make our
notation more compact.}.

We can also repeat the t-channel computation in this
symmetric formalism
\begin{equation}
A \sim \left \langle {\cal V}^{T_c} \right\rangle_{D_2} \ \Pi^c(\apr t)
\ \left \langle {\cal V}^{T_c} \ V^T_1 \ V^T_2 \right \rangle_{S^2}
\end{equation}
A short computation of the two individual diagrams and contraction
using the propagator results in the expression (\ref{2Tpomcl}).

At this point we should make two comments. First, the reason we
have insisted on this symmetric formulation is that although in
the present paper we will discuss two-particle BCFW deformation of
the amplitudes there exist more general ones involving three or
more deformed momenta. These have appeared in the literature in
various works \cite{Risager:2005vk, Benincasa:2007xk}. The behavior
of amplitudes under more general deformations can be studied using
this symmetric formalism by grouping two, three or more particles
on the left part of the amplitude and the rest on the right one.
We will say a bit more on this point in the conclusions.

Second,
note that for the corresponding superstring scattering of level
one states, that is the graviton, the  dilaton and the Kalb--Ramond field
\cite{Hashimoto:1996bf}, the two pomeron channels correspond to
the field theory channels which reproduce the expected behavior
from the gravity plus DBI actions (see for example
\cite{Garousi:1998fg}). From a technical point of view this comes
from the fact that in these amplitudes there is an overall beta
function expression which multiplies an expression which does not
depend on  $\apr$ and therefore it goes into itself in the field
theory limit $\apr \to 0$. The field theory  poles come from
massless poles of the beta function expression. When we take the
opposite limit, i.e.~Regge one, the pomerons effectively
average over the zeros and poles of the beta function which lie on
the real axis of the kinematic variable \cite{Green:1987sp}. It is
therefore no surprise that the structure and factorization
channels of the two different limits agree. But it is highly
non-trivial in more complicated amplitudes where this simple
structure is not expected to persist.

\subsubsection{Application 1: closed string tachyon scattering with
three open strings}\label{PomEx}

In this section we would like to demonstrate how the pomeron operators
can be used to extract the Regge behavior for amplitudes involving
both closed and open string states. Our results will be relevant for the comparison with
the field theory expectations in the spirit of
\cite{ArkaniHamed:2008yf}. In principle we should consider  an
amplitude closely related to  (\ref{2TD}). It has been
shown in  \cite{Kawai:1985xq} that closed string amplitudes can
be constructed by gluing of open string amplitudes.
The four open
string case on the disc can be used in similar manner to construct
\cite{Garousi:1996ad} the two closed and the two open - one closed
string amplitudes based on the identification (\ref{substitute}).  
Recently this idea has been pushed even further
 to demonstrate in full generality the
relation of pure open string
amplitudes to amplitudes with both open and closed strings 
 \cite{Stieberger:2009hq}.

In the previous section we studied the two closed tachyons
amplitude. The next case one could try is amplitude of two open tachyons with momenta $k_1, k_2$
and one closed tachyon with momentum $p$
\begin{equation}\label{2t1T}
{\cal A}(p,k_1,k_2) \sim {\Gamma(1-\apr t) \over \Gamma\left(1-\hapr t\right)^2}\ .
\end{equation}
Due to momentum conservation there is only one kinematic
variable $t=-2 k_1 \cdot k_2= 2k_1 \cdot p+\frac{2}{\alpha'}=2 k_2
\cdot p+\frac{2}{\alpha'}$ and the amplitude does not  exhibit
Regge type behavior. One can indeed see that its  large $t$
behavior is not a pomeron like one. This will actually be
important for our discussion in section \ref{FT}.

The next case to consider is the three open and one closed tachyon
amplitude. The corresponding amplitude in the superstring for
three scalars and one graviton has been computed in
\cite{Fotopoulos:2001pt}. The bosonic tachyon case can be
extracted easily
\begin{eqnarray}\label{3t1T}
{\cal A}(p,k_1,k_2,k_3) \displaystyle &\sim &  \int_{\cal{H}^{+}} d^2 \!z \;
|1-z|^{\apr (s+u)/2 -2} |z|^{\apr (t+u)/2-2} (z-\bz)^{-\apr(s+t+u)/2+1}   \nonumber \\
& \sim & { \Gamma\left({1\over 2}-\apr {s  \over 4}\right)\  \Gamma\left({1\over
2}-\apr {t \over 4}\right)\ \Gamma\left({1\over 2}-\apr {u \over 4}\right) \
\Gamma\left(1- \apr {s+t+u \over 4}\right) \over \Gamma\left(1-\apr {s+u \over
4}\right)\ \Gamma\left(1-\apr {s+t \over 4}\right)\ \Gamma\left(1-\apr {t+u \over 4}\right)}
\end{eqnarray}
where the kinematic variables are defined as
\begin{equation}\label{kin3t1T}
s=-4 k_1 \cdot k_2\ , \quad t =-4 k_1 \cdot k_3\ , \quad u =-4 k_2 \cdot
k_3 \end{equation} and they satisfy
\begin{equation}\label{kin3t1Tb}
2 (V \cdot p)^2 +s+t+u={6 \over \apr} \ .
\end{equation}

First lets try to find a Regge regime of the s-channel type as in
(\ref{s-int}). This corresponds to the factorization of the three
open string states on a disc and the closed string state on
another disc connected with a strip (open string state). But as in
the case of (\ref{2t1T}) there is no Regge type behavior with such
a factorization. This is due to momentum conservation
\begin{equation}\label{kin3t1Tc}
V\cdot p+k_1+k_2+k_3=0
\end{equation}
which dictates that if we highly boost the three open string states
in one direction the closed string is boosted as well and
there is no pomeron exchange among them.

In the supersymmetric computation of \cite{Fotopoulos:2001pt} the
amplitude is the product of a an expression similar to
(\ref{3t1T}) and a kinematic prefactor which depends on the
momenta and polarizations of the scattered particles. In that case
the amplitude has a factor  $\Gamma\left(- \apr{s+t+u \over 4}\right)$
whose massless pole give rise to the field theory limit of the
amplitude. From the discussion above it should be clear that there
is no Regge type behavior which leads to the same factorization on
the world-sheet as the field theory limit does. This will be
important in our discussion in section \ref{FT}.

Instead, we can
choose to boost only two open string tachyons, say those with momenta $k_1$ and $k_2$, which
corresponds to the kinematic region
\begin{equation}\label{Reggekin3t1T}
(V\cdot p)^2, s << \hht , \hat{u}\ .
\end{equation}
In this limit
\begin{equation}
\frac{\Gamma\left(\frac{1}{2}-\alpha'\frac{u}{4}\right)}{\Gamma\left(1-\alpha' \frac{s+u}{4}\right)} \sim
\left(\frac{\alpha' \hat u}{4}\right)^{\frac{\alpha' s}{4}-\frac{1}{2}} e^{-\frac{\alpha' s}{4}}
\end{equation}
and similarly for the ratio of the other two Gamma function involving $\hat t$ and $s$. Since
 $\hat t \sim -\hat u$ due to (\ref{kin3t1Tb}) we finally obtain
\begin{equation}\label{Regge3t1T}
{\cal A}(p, \hat k_1, \hat k_2,k_3) \sim e^{-\frac{\alpha' s}{2}}
\left({\alpha'\hat{u}\over 4}\right)^{\frac{\alpha' s}{2}-1}
\Gamma\left({2-\apr s\over 4}\right) {\Gamma\left( \frac{\apr
(V\cdot p)^2-1}{2}\right) \over \Gamma\left( \frac{\alpha' (V\cdot
p)^2-1}{2}+\frac{\alpha' s}{4}\right)}\ .
\end{equation}

Let us know see how we can reproduce  this behavior  in terms of the pomeron.
Using the open pomeron vertex of
\cite{Cheung:2010vn}\footnote{Notice that $\alpha'_{here} = 2 \alpha'_{there}$.}
 we arrive at the expression
\begin{eqnarray}\label{Pom3t1T}
{\cal A}(p, \hat k_1, \hat k_2,k_3)  &\sim& \int_{0}^{\infty} dx
\left \langle \big(c(x)+c(-x)\big) \ c(i)
\ \tilde{c}(-i)\right\rangle \times \Gamma\left(1-\hapr s\right) \times \nonumber \\
 && \left\langle \Big((\hk_1-\hk_2)\cdot \partial X\Big)^{\hapr s-1}
e^{2i (\hk_1+\hk_2) \cdot X} (x)\  e^{2i k_3 \cdot
X}(-x) \ e^{i p \cdot X}(i) \ e^{i D p \cdot X}(-i)\right\rangle \ . \nonumber \\
 \end{eqnarray}
This amplitude contains the open pomeron corresponding to the highly boosted
 strings with momenta $\hat k_1$ and $\hat k_2$ and the unboosted  open and closed tachyons
 with momenta $k_3$ and $p$ respectively. We have inserted the open pomeron at $x$ on
 the boundary, the open tachyon at $-x$ and the closed tachyon at $(z,\bar z)=(i,-i)$.

 After performing  the operator contractions we obtain
 \begin{equation}
 {\cal A}(p, \hat k_1, \hat k_2,k_3)  \sim   \big(\alpha'(\hat u- \hat t) \big)^{\frac{\alpha' s}{2}-1}
 \Gamma\left(1-\hapr s\right)
 {\cal I}(s, V\cdot p)
\end{equation}
 where the integral
\begin{eqnarray}\label{int3t1T}
 {\cal I}(s, V\cdot p) = \int_{0}^{\infty}  d y \
(y-1)^{\apr s /2 -1} y^{(\apr (V\cdot p)^2 - 3)/2} (1+y)^{-\apr
(V\cdot p)^2 + 2 -\apr s /2}= &&\nonumber \\
{\Gamma ^2\left(\hapr (V\cdot p)^2-{1\over 2}\right) \over \Gamma\left(\apr(V \cdot p)^2-1\right)} \
 _2F_{1} \left(1-\hapr s,\hapr (V \cdot p)^2-{1\over 2}; \apr (V \cdot p)^2-1;
2\right) \end{eqnarray} can be found in \cite{Grads}.
 Using the
 identities \cite{Bateman}
 \begin{eqnarray}\label{hyper}
_2F_{1} \left(1-\hapr s,\hapr (V \cdot p)^2-{1\over 2}; \apr(V\cdot
p)^2-1; 2\right)=  \nonumber \\
  (-1)^{\apr s/4 -1/2} \;
 _2F_{1} \left( \hapr
s-{1\over 2},\hapr (V \cdot p)^2+{\apr s\over 4}-1; \hapr (V\cdot
p)^2; 1\right)
\end{eqnarray}
and
\begin{equation}
_2F_{1}(a,b;c;1)=\frac{ \Gamma(c) \Gamma(c-a-b)}{  \Gamma(c-a) \Gamma(c-b)}
\end{equation}
 we can indeed match this result with the expression in
(\ref{Regge3t1T}). This verifies our formulation and moreover
demonstrates that Regge behavior is indeed distinct from the field
theory limit making the conjecture of eikonal Regge region of
\cite{Cheung:2010vn} non-trivial.

We would like to make a final comment before moving on to the next
section. In (\ref{Pom3t1T}) we chose  to boost the momenta $\hk_1,
\hk_2$ in order to compute the amplitude. We therefore used the
pomeron vertex form for the OPE of two open string tachyons. We
could have chosen to work with the OPE of the closed string with
one open string, corresponding to a boost of $\hp$ and, say,
$\hk_3$. This leads to a new pomeron,  ${\cal V}^{T_c T_o}$,
which describes the mixing of one open with one closed string tachyon.
The OPE is in principle more complicated since
it has two singularities to integrate over, one from the closed
string operator approaching the open string and a second one from
the self-interaction of the closed string operator with its image
when it approaches the boundary. Nevertheless, if
$\apr \hp\cdot D\cdot \hp=\apr \hp\cdot V \cdot \hp=4$, that is when the
closed string has only world-volume momenta, the computation using
the OPE is rather easy and it leads to the pomeron operator discussed  in
appendix B. One then proceeds with the three point amplitude
of this pomeron operator and the other two open string tachyons on
the disc reproducing once more the limiting behavior
(\ref{Regge3t1T}).

\subsubsection{Application 2:  ultrarelativistic D-brane scattering}\label{ultra}

In this section we will compute the absorptive part of the 1-loop
partition function of two ultrarelativistic branes using
pomeron operators. Our result will be
compared to the result of \cite{Bachas:1995kx}.

We would like to consider scattering of D-branes which move at
high relative velocities very close to the speed of light. The
scattering of two such objects is dictated, in the lowest order in
perturbation theory, by computing the one-loop partition function
for the open strings stretching between them. Through world-sheet
duality this is related to the emission of a closed string state
from one brane and absorption by the other. We will concentrate to
the bosonic case although the supersymmetric one is totally
analogous.

Lets begin by defining the setup and a few useful quantities. For
a D$p$-brane moving in a $d$-dimensional spacetime
with velocity $\vec{V_1}$ we have the momentum
vector
\begin{equation}
P_1^\mu=M_p \ \left({1\over \sqrt{1-V_1^2}},
{\vec{V_1}\over \sqrt{1-V_1^2}}\right)
\end{equation}
where $M_p$ is the
mass of the D-brane. Assume that the two D-branes have velocities
very close to the speed of light, $|\vec{V_1}|=|\vec{V_2}|=1-q, \
q \ll 1$, and opposite to each other $\vec{V_1}=-\vec{V_2}$. The
kinematic invariant describing the scattering process is
\begin{equation}\label{s}
s= -P_1\cdot P_2= M_p^2 \left( {1- V_1 V_2  \over
\sqrt{1-V_1^2}\sqrt{1-V_2^2}} \right) \simeq M_p^2 {1\over q}\ .
\end{equation}
The D-branes have world-volume coordinates $x^\a,\ \a =0,\dots, p$
with Neumann boundary conditions while the transverse directions
 $x^i,\ i =p+1,\dots, d-1$ have
Dirichlet.
For simplicity assume that the two D-branes
move oppositely along the direction $x^{d-1}$ and are separated by
a distance $b$ along the direction $x_{\perp}=x^{d-2}$. We define
$x^{\pm} = {1\over \sqrt{2}}(x^0 \pm x^{d-1})$ light cone
coordinates. The first brane is boosted along the $x^+$ direction
and the second one along $x^-$.

The pomeron computation for this scattering takes the form\footnote{
 Notice that here $ {\cal V}^{T_c T_c}$ is the stripped pomeron
vertex, see also footnote 7.}
\begin{eqnarray}\label{branescatt}
{\cal A}&& \sim  \left\langle B,V_1| \  \Delta \ |B,V_2\right\rangle  \simeq
\nonumber \\
&&\simeq \ _{(D_2,V_1)}\left \langle {\cal V}^{T_c T_c}(0) \right\rangle \
\Pi^c(\apr \vec{k}^2_\perp)\left \langle {\cal V}^{T_c T_c}(0)
\right\rangle_{(D_2,V_2)}
\end{eqnarray}
where $\Delta$ is the closed string propagator
and $|B,V\rangle$ is  the boundary state for a brane
moving with velocity $V$.
 In the second line we use the CFT
approach and consider the boundary
conditions on the disc $D_2$ for moving branes with
velocities $V_1$ and $V_2$, while
$\vec{k}_\perp$ is  the  pomeron momentum which is transverse  to the D-branes and to their line of
motion, i.e.~it is along the $x_{\perp}$ direction. In the notation
of subsection 3.1
we have
$-t+8/\apr=p^2=\vec{k}_\perp^2$. The pomeron propagator
(\ref{tPP}) takes
explicitly the form
\begin{equation}\label{tPP2}
\Pi^c(\apr \vec{k}^2_\perp )= {\Gamma\left(-1+{\apr
\vec{k}^2_\perp \over 4}\right)\over \Gamma\left(2-{\apr
\vec{k}^2_\perp \over 4}\right)} e^{i\pi \left(-1+{\apr \vec{k}^2_\perp
\over 4}\right)}.
\end{equation}

We use now  the one-point function of the pomeron operators (\ref{tPom2})
on the D-brane to compute \begin{equation}\label{onepoint} \left \langle
{\cal V}_{T_c T_c}(0) \right\rangle_{(D_2,V_1)} \sim \left\langle
:\left({\partial X^+ \over \sqrt\apr}\right)^{1- \apr
\vec{k}_\perp^2/4} e^{i p\cdot X}:\ :\left({\bpartial \bar X^+ \over
\sqrt\apr}\right)^{1- \apr \vec{k}_\perp^2/4} e^{i p\cdot
\bar X}: \right\rangle_{(D_2, V_1)}\ .
\end{equation}
 We have assumed that $V_1$ is
along $x^+$ and inserted the appropriate pomeron vertex normalized
as in \cite{Brower:2006ea}. The correlator is computed on the
boundary state of a boosted D-brane \cite{Billo:1997eg}.
One can compute the matrix $M(V)$ defined in \cite{Billo:1997eg},
which describes the gluing of holomorphic and antiholomorphic
fields on the moving D-brane, in the ultrarelativistic limit. It
turns out that the leading correlator, on the disc, is
\begin{equation}\label{corrmov}
X^{+}(z) \bar X^+ (\bw)\sim -\hapr {1\over q} \ln(1-z\bw)\ .
\end{equation}
The brane glues together the holomorphic $X^+$ and antiholomorphic
$\bar X^+$ string coordinates. Notice that for light cone coordinates the only non-vanishing correlators between
holomorphic fields is $\left \langle X^+(z) X^-(w)\right \rangle $ and
similarly for the antiholomorphic fields.

We will also need the
boundary state normalization from \cite{Billo:1997eg}
\begin{equation}\label{branenorm}
{\cal N}_{V}=\left \langle 1 \right\rangle_{(D_2, V)} \simeq (2\pi
\sqrt{\apr})^{d_\perp} \sqrt{1-V^2} \; \delta (x^{d-1}-x^0 V-
y^{d-1}) \prod_{i \neq d-1} \delta(x^i-y^i)
\end{equation}
where we have defined $d_\perp=d-p-2$, ignored factors of $\pi$
and kept only factors depending on $\apr$ and $V$. The delta function
describes the motion of the brane in spacetime and depends on the
zero modes $x^i, \ i=p+1,\dots, d-1$ and the initial position
$y^i$. The final result, after combining all the ingredients and
using  Euler's reflection formula $\Gamma(x) \Gamma(1-x) = \displaystyle \frac{\pi}{\sin \pi x}$,
 takes the form
\begin{equation}\label{branescatt2}
{\cal A} \sim V_p {(2\pi \sqrt\apr)^{2d_\perp}\over |V_1-V_2|}
\sqrt{1-V_1^2}\sqrt{1-V_2^2} \left( {s\over M_p^2} \right)^{2-\apr
\vec{k}_\perp^2/2} F\left(\frac{\apr\vec{k}_\perp^2}{2}\right)
\end{equation}
where
\begin{equation}\label{F}
F\left(\frac{\apr\vec{k}_\perp^2}{2}\right)=
- \frac{e^{i\pi \left(-1+{\apr \vec{k}^2_\perp \over 4}\right)} }{\sin \pi
\left(-1+{\apr \vec{k}^2_\perp \over 4}\right)}\ .
 \end{equation}
 The factor
$|V_1-V_2|$ in the denominator of (\ref{branescatt2})  comes
from integration over the zero modes of the two delta functions of
the boundary states in (\ref{branenorm}). In the
ultrarelativistic limit under consideration we have
\begin{equation}\label{ultraident}
{\sqrt{1-V_1^2}\sqrt{1-V_2^2}\over |V_1-V_2|} \simeq q \simeq
{M_p^2 \over s}\ .
\end{equation}

 One can alternatively boost the branes back to their
approximate rest frames extracting the boost parameter $(s/s_0)$
where $s_0=2M_p^2$ is the rest frame center of mass energy. Then, as
explained in \cite{Brower:2006ea}, the equation (\ref{branescatt})
becomes
\begin{equation}\label{branescattBPST}
{\cal A} \sim  {\cal N}_{V_1}{\cal N}_{V_2} \left({s\over s_0}\right)^{2- \apr
\vec{k}_\perp^2/2}\left \langle {\cal V}^{T_c T_c}(0)
\right \rangle_{(D_2,V_1)} \ \Pi^c(\apr \vec{k}^2_\perp)\left \langle {\cal
V}^{T_c T_c}(0) \right \rangle_{(D_2,V_2)}\ .
\end{equation}
Notice that compared to (3.20) of \cite{Brower:2006ea}, the
normalization factors (\ref{branenorm}) need to be taken into
account in order to derive the correct result.

Let us finally compute absorptive part of the scattering
amplitude, ${\rm Im}(\delta)$,  in order to compare with the
corresponding expression in \cite{Bachas:1995kx}. This is given by
the Fourier transform in position space of the imaginary part of
the expression in (\ref{branescatt2}). Notice that from (\ref{F})
${\rm Im}(F)=-1$. We can easily show, as in \cite{Brower:2006ea} and
using (\ref{ultraident}), that
\begin{eqnarray}\label{branescattabs}
{\rm Im}(\delta) && \simeq  \int d^{d_\perp} k_\perp \ e^{i
\vec{k}_\perp \cdot \vec{x}_\perp} {\rm Im ({\cal A})} \simeq \nonumber \\
&&\simeq V_p (2\pi \sqrt{\apr})^{2d_\perp} \left( {s\over M_p^2}
\right)\int d^{d_\perp} k_\perp \ e^{i \vec{k}_\perp \cdot
\vec{x}_\perp} e^{-\hapr \vec{k}_\perp^2 \log\left({s\over
M_p^2}\right)}\simeq \nonumber \\
&&\simeq V_p \ {s\over M_p^2}\left[\log\left({s\over
M_p^2}\right)\right]^{-{d_\perp \over 2}} \ e^{- {b^2 \over 2\apr
\log\left({s\over M_p^2}\right)}}
\end{eqnarray}
with the impact parameter $x_\perp=b$.
This is indeed the result of \cite{Bachas:1995kx} for
$d=10$. The
scattering has the diffusive form in transverse position space
typical of Regge scattering. The characteristic size, as seen form
an observer placed on the other fast moving brane,  grows with the
center of mass energy as $  b^2  \sim \apr
\log \left({s\over M_p^2}\right)$ .

\subsection{Pomeron operators for level one states}\

In this subsection we give the pomerons for
level one states in the bosonic and superstring theories. These
operators can be easily derived using similar OPE techniques as
we used earlier for the tachyons. In all cases we kept the leading and sub-leading terms
in the OPEs before integrating.  Some of the results are given in
\cite{Cheung:2010vn, Boels:2010bv} and we list them along with our results.

\subsubsection{Bosonic pomerons}

The
bosonic operators are as follows. The one involving two gauge bosons (or transverse scalars) reads
\begin{eqnarray}\label{opbospom0}
{\cal V}_{bos}^{AA} &\sim & \he_1^\mu \he_2^\nu \ e^{2i k \cdot X}
\Gamma(2\apr k_1\cdot k_2) (-2 i  Q \cdot
\partial X)^{-2\apr k_1 \cdot k_2} \nonumber \\ && \left(  {-2 i Q \cdot \partial X \over
 2\apr k_1\cdot k_2-1} ( 2\apr k_\mu k_\nu-\eta_{\mu \nu} ) +
2i (\partial X_\mu k_\nu - \partial X_\nu k_\mu) \right)
\end{eqnarray}
where $k=\hk_1 +\hk_2= k_1+k_2$ and $2Q= \hk_1-\hk_2$\ .

The pomeron with two massless closed strings (gravitons,
antisymmetric tensors or dilatons) in the bulk is
\begin{eqnarray}\label{bospomcl}
{\cal V}_{bos}^{GG}&\sim&- \he_1^{\mu \bmu} \ \he_2^{\nu \bnu} \
e^{i p \cdot X}\ \Pi^c(\apr  p_1\cdot p_2)
  (-Q \cdot
\partial X \ Q\cdot \bpartial X)^{-\hapr p_1 \cdot p_2} \\
&& \left[ (Q\cdot \partial X) P_{\mu \nu} - K_{\mu \nu} \left(
\hapr p_1\cdot p_2 -1 \right) \right]  \left[ (Q \cdot \bpartial
X) \bar{P}_{\bmu \bnu} - \bar{K}_{\bmu \bnu} \left( \hapr p_1\cdot
p_2-1\right) \right]\nonumber
\end{eqnarray} where  $p=\hp_1+\hp_2,\
2Q=\hp_1-\hp_2$ and
\begin{equation}\label{defPK}
P_{\mu \nu}=  \hapr p_\mu p_\nu -\eta_{\mu \nu} \ , \qquad K_{\mu
\nu}=\partial X_\mu p_\nu - \partial X_\nu p_\mu \ ,
\end{equation}
\begin{equation}
\Pi^c(\apr  p_1\cdot p_2) = {\Gamma\left( \hapr p_1\cdot
p_2- 1\right)\over \Gamma\left(2-\hapr p_1\cdot p_2\right)} \
e^{i\pi \left(-1+ \hapr p_1\cdot p_2\right)}\end{equation}
  The form of the operator is
determined either from the previous pomeron vertex
by KLT or
explicitly by computing the
OPE of two operators at positions $w_1$ and $w_2$ and then
integrating over their  relative position $ (w_1-w_2)/ 2$ as in
the tachyon case (\ref{tOPE})-(\ref{tPom1}) (see Appendix A for the
explicit derivation). We use barred and unbarred indices to
distinguish the two independent Lorentz spin symmetries in  the
spirit of \cite{ArkaniHamed:2008yf}.

Finally, for the pomeron involving a massless closed string state
on the disk (tadpole) we can use the doubling trick
\cite{Garousi:1996ad} which allows to construct a graviton
vertex operator as a direct product of two gauge boson operators. The
polarization is given by
\begin{equation}\label{dtrick}
\e_{\mu \bar{\lambda}} D^{\bar{\lambda}}_\nu=(\e D)_{\mu \nu} =
\e_\mu \otimes \e '_\nu\ .
\end{equation}
Of course the pomeron operator can be constructed directly using
similar steps as in appendix A. The corresponding vertex is given
by the expression
\begin{eqnarray}\label{bospomtad}
&&{\cal V}^{G}_{bos} \sim  \he^{\mu \bnu} \ e^{2i V\cdot \hp \cdot
X} \Gamma\left(-1+\hapr \hp\cdot D\cdot \hp\right) (-i Q \cdot
\partial X)^{-\hapr \hp\cdot D\cdot \hp} \Bigg[ (-i Q \cdot \partial X )(D\cdot P)_{\mu \bnu}
 \nonumber \\
&&+ 2i  \Big(\partial X_\mu (V\cdot \hp)_{\bnu} - \partial (D\cdot
X)_{\bnu} (V\cdot \hp)_\mu\Big) \left(-1+\hapr \hp\cdot D\cdot \hp\right)  \Bigg]
 \end{eqnarray} where $2Q= \hp-D\cdot \hp=
2 N\cdot \hp$, we have used the fact that $\he^{\mu \bar \nu}
D\cdot \hat{ p}_\nu= 2 \he ^{\mu \nu} (V\cdot \hp)_{\bar \nu}$ due
to the physical state condition and defined $P_{\mu
\nu}=-\eta_{\mu \nu}  + 2\apr (V\cdot \hp)_\mu (V\cdot \hp)_{
\nu}$. Notice the difference of the prefactors compared to the
analogous result  for the Type I superstring in the zero picture
in \cite{Cheung:2010vn}. Our prefactors have tachyon poles as
expected for the bosonic theory.

\subsubsection{Superstring pomerons}

For the
type II superstring with D-branes, which is the case of interest
in the present study, the zero picture results can be found in
\cite{Cheung:2010vn}.

For two gauge bosons both in the -1 picture, we derive the following
\begin{eqnarray}\label{opsusypom0}
{\cal V}_{susy}^{A(-1)A(-1)}  &\sim&  \he_1^\mu \he_2^\nu \ e^{-2\varphi}
 e^{2i k \cdot
X} \Gamma(2\apr k_1\cdot k_2 -1) (-2 i  Q \cdot
\partial X)^{-2\apr k_1 \cdot k_2}
 \nonumber \\
&&\Big(\eta_{\mu \nu}(-2 iQ\cdot
\partial X)
- \psi_\mu \psi_\nu
(2\apr k_1\cdot k_2 -1)  \Big)
\end{eqnarray}
where we have designated the superghost charge of the operator.
As in the previous examples we have defined  $k=\hk_1 +\hk_2= k_1+k_2$ and $2Q= \hk_1-\hk_2$\ .

For two massless closed strings we choose an asymmetric picture
for the graviton operators. The reason for this is that if we wish
to insert them on the disc we should not exceed the total
superghost charge of $-2$, otherwise we will have to insert
picture changing operators. We will compute the OPE of
$V_{(0,-1)}$ with $V_{(-1,0)}$ by calculating separately the
holomorphic OPE and using KLT to derive the closed string OPE. The
holomorphic OPE we wish to compute is
\begin{equation}\label{susyholpomcl}
V_{(0)}(y) V_{(-1)}(-y) \sim \he_{1;\mu} \Big(\partial X^\mu +i\apr
(\hk_1 \cdot \psi) \psi^\mu\Big) e^{2i \hk_1\cdot X}(y) \;\; \he_{2;\nu}
\psi^\nu e^{-\varphi} e^{2i \hk_2 \cdot X}(-y)\ .
\end{equation}
The final result (see appendix A for some details)
after the integration over
$y$ is
\begin{eqnarray}\label{susyholpomcl2}
&&{\cal V}^{A(-1)A(0)}_{susy}  \sim \he_{1;\mu} \he_{2;\nu} \ e^{-\varphi}
\ e^{2i k
\cdot X} \Gamma(2\apr k_1\cdot k_2-1) (Q\cdot
\partial X)^{-1-2\apr k_1\cdot k_2} \nonumber \\
&&
\Big[ \eta^{\mu\nu}(Q\cdot
\partial X)(\hk_1 \cdot \psi)
+
 (2 \apr k_1\cdot k_2 -1) \times  \\
 &&
 \left( -2(\hk_1 \cdot \psi) \psi^\mu
\psi^\nu + \eta^{\mu\nu} \left( (\hk_1 \cdot \psi) \dot{\phi}+
(\hk_1\cdot \dot{\psi}) \right) \right) - (Q\cdot
\partial X)( \psi^\mu k^\nu-\psi^\nu k^\mu )  \Big]\ . \nonumber
\end{eqnarray}

Now we can take the product of two such
pomeron operators, one of which corresponds to the OPE of the
holomorphic parts of the operators and the other to the
antiholomorphic,  to obtain the closed string pomeron. One only needs
to make the substitutions $2k_i \to p_i$ for the holomorphic part,
$2k_i \to D\cdot p_i$ for the antiholomorphic and to use the
polarizations map $\he_{i;\mu} \times \he_{i;\nu} \to (\epsilon
 D)_{\mu \nu}= \epsilon_{\mu \bmu} D^{\bmu}_\nu$.  Moreover, we get a product of two gamma functions
 which can be manipulated to give the ratio of two gamma functions which forms part of the closed pomeron propagator
 $\Pi^c(\apr s)$ of \cite{Brower:2006ea}. The only difference is that we get an extra
 trigonometric phase
 factor of the kinematic invariant. This of course can be attributed to the KLT relations which generically require such
 factors when gluing open string amplitudes to construct closed string ones \cite{Kawai:1985xq}.
 The final result is rather long and not very illuminating, therefore we present only the leading behavior
\begin{eqnarray}\label{susypomcl0}
 {\cal V}_{susy}^{GG}&=& -{\rm tr}(\he_1\cdot
\ \he_2) \ e^{-\varphi -\bar \varphi} e^{i p \cdot X} \ \Pi^c(\apr  p_1\cdot
p_2) \
  (-Q \cdot
\partial X \ Q\cdot \bpartial X)^{1-\hapr p_1 \cdot p_2}\ .
\end{eqnarray}

Finally the pomeron corresponding to the tadpole diagram is
\begin{eqnarray}\label{susypomtad}
{\cal V}^{G}_{susy}  &\sim&  \he_{\mu \bnu}\
e^{-\varphi -\bar \varphi}
e^{2i V\cdot p \cdot X}
\Gamma\left(-1+\hapr p\cdot D\cdot p\right) (-i
Q \cdot \partial X)^{-\hapr p\cdot D\cdot p} \nonumber \\
&&\left( D^{\mu \bnu}(-i Q \cdot \partial X) - \psi^\mu (D\cdot
\psi)^{\bnu} \left(-1+\hapr p\cdot D\cdot p\right) + \dots \right)
\ ,
\end{eqnarray}
where $2Q= \hp-D\cdot \hp= 2 N\cdot \hp$.

The results above for the superstring have tachyonic poles, just
like the bosonic ones, as commented in \cite{Cheung:2010vn}, which
should be cancelled by the OPE with vertices in the rest of the
amplitude. These poles would not have been present had we computed
the pomeron operators in the 0 picture. Notice that the tadpole
operators above have a prefactor typical of an open string
pomeron as expected. Remember though that they correspond to the
OPE of a closed string operator with itself due to the non-trivial
boundary conditions.

\section{BCFW  shifts in the presence of  defects}\label{BCFWrev}

\subsection{A short review of BCFW}

The key point of BCFW \cite{Britto:2005fq} is that tree
level amplitudes constructed using Feynman rules are rational
functions of external momenta. Analytic continuation of these momenta on
the complex domain  turns the amplitudes
into meromorphic functions
which can be constructed solely by their residues. Since the latter are
products of lower point on-shell  amplitudes the final outcome is a set
of powerful recursive relations.

The simplest complex deformation involves only two external particles whose momenta are shifted as
\begin{eqnarray}\label{BCFWshift1}
 \hp_1(z)= p_1 + q z\ , \quad
 \hp_2(z)= p_2 - q z \ .
\end{eqnarray}
Here $z \in \mathbb{C}$ and to keep the on-shell condition we  need
$q \cdot p_{1} =q\cdot p_2= 0$ and  $q^2 = 0$. In Minkowski spacetime this is only possible for
complex $q$. In particular, if the dimensionality of spacetime is  $d \geq 4$,
we can choose a reference frame where
the two external momenta $p_1$ and $ p_2$ are back to back with equal
energy scaled to 1 \cite{ArkaniHamed:2008yf} \be\label{mom} p_1  =
(1,1,0,0,\ldots,0), \quad p_2 = (1,-1,0,0,\ldots,0), \quad q = (0,0,1,i,0,\ldots,0)\ .
\ee
For gauge bosons the polarizations under a
suitable gauge can be chosen as \be\label{gpol} \epsilon_1^- =
\epsilon_2^+ = q, \quad \epsilon_1^+ = \epsilon_2^- = q^*, \quad
\epsilon_T = (0,0,0,0,\ldots,0,1,0,\ldots,0) \ .
\ee
Under the deformation
(\ref{BCFWshift1}) the polarizations become
 \begin{equation}
 \label{gpolshift} \he_1^-(z) =
\he_2^+(z) = q, \quad \he_1^+(z) = q^* + z p_2, \quad \he_2^-(z) = q^*
- z p_1, \quad
\he_T(z) = \epsilon_T
\end{equation}
so that they are orthogonal to the shifted momenta.

A general amplitude, which after the deformation becomes a meromorphic function
 ${\cal A}_n(z)$,
 will have simple poles for those values of $z$
where the propagators of intermediate states go on shell
(on the complex plane)
\be\label{BCFWpole} \frac{1}{P_J(z)^2} =
\frac{1}{P_J(0)^2 - 2 z q \cdot P_J}\ .
 \ee
 The undeformed amplitude
can be computed using Cauchy's theorem:
\begin{equation}\label{BCFWrel1}
{\cal A}_n (0) = \oint_{z=0} \frac{{\cal A}_n(z)}{z} dz = - \left\{\sum
\mathrm{Res}_{z=\textrm{finite}} + \mathrm{Res}_{z= \infty}
\right\} \ .
\end{equation}

As already stated the residues at finite locations on the complex
plane are necessarily, due to unitarity, products of lower point
tree level amplitudes alas computed at complex on-shell momenta.
The residue at infinity can have a  similar interpretation is some
special cases \cite{Feng:2009ei,Feng:2010ku} but in general
cannot be written as product of lower point amplitudes. In most
cases involving gauge bosons and/or gravitons there is an
appropriate choice of the deformed external polarizations such
that under a shift of the type (\ref{BCFWshift1}), ${\cal A}_n(z)$
vanishes in the limit $z \rightarrow \infty$. In these cases the
BCFW relation takes the simple form
\begin{equation}\label{BCFWrel2}
{\cal A}_n(1,2,3 \ldots, n) = \sum_{r,h(r)} \sum_{k=2}^{n-2}
\frac{{\cal A}_{k+1}(1,2,\ldots, \hat{i}, \ldots, k, \hat{P}_r)
{\cal A}_{n-k+1}(\hat{P}_r, k+1, \ldots, \hat{j}, \ldots, n)}{\left(p_1 +
p_2 + \ldots + p_k \right)^2 +  m_r^2} \ ,
\end{equation}
where hat is used for the variables which are computed at the
residue of the corresponding pole (\ref{BCFWpole}) and satisfy the
physical state condition.  Since we have made a complex
deformation these are complex momenta unlike the momenta of the
rest of the external particles. The summation is over all particle
states in the spectrum of the given spectrum and their
polarization states.

As was shown in \cite{Cheung:2010vn, Boels:2010bv}, tree level string amplitudes
for appropriate regimes of the Mandelstam variables
 can be made to vanish when $z \to \infty$. Therefore one expects that these amplitudes
 satisfy also
 recursive relations similar to the field theoretic ones (\ref{BCFWrel2}).
 After this very
short review of BCFW we will proceed by considering BCFW shifts
in the existence of spacetime defects.
The discussion that
follows is general and does not confine solely to strings. It
could also possibly be useful for cases of macroscopic defects like
p-branes or those appearing in brane-world models.

\subsection{BCFW shifts in the presence of defects}\label{BCFWdef}

In the presence of a defect (brane)
 the fact that momentum is not conserved in the
directions normal to the brane\footnote{We mean hard (i.e.~shifted) momentum
conservation. Soft momenta can be non-conserved without affecting
the analysis.} implies that there are several novelties when considering
BCFW shifts.
Let us  consider the scattering of a number $m$ of
 bulk states carrying momenta $p_J, J=1,\ldots,m$ off
 a brane. In string theory, for instance, this is the case where a number
 of closed string states are scattered on a D-brane. The generalization
 where both bulk and brane states (closed and open strings
 respectively in the D-brane case) are scattered  is straightforward.
We define the kinematic variables
\begin{equation} \label{stvar}
s_{ij}= 2 p_i \cdot V\cdot p_j\ , \quad t_{ij}=-2 p_i  \cdot p_j
\end{equation} which can be used to parametrize a general
scattering process involving branes. Recall that $V$ is the
matrix that projects momenta along the world-volume directions
of the brane.
Due to momentum conservation
along the brane, which reads
\begin{equation}
\sum_{j=1}^m V \cdot p_J =0 \ ,
\end{equation}
 the aforementioned invariants are not all
independent. Moreover, if the momentum of a particle, say $p_k$, is
along the brane obviously  $t_{ik}=-s_{ik}$.
As in standard BCFW there is a plethora of multi-particle shifts
\cite{Risager:2005vk} but we will be interested only in the cases
where either one or two momenta are shifted.

\paragraph{One momentum shift.} The ability to shift
only one momentum without world-volume violating momentum
conservation is a novel feature allowed by the defect. For
instance, we can shift the momentum of particle 1 as
\begin{equation}
\hat p_1(z) = p_1 +z q
\end{equation}
with the on-shell condition and momentum conservation dictating
\begin{equation}
\label{oneparshi}
p_1 \cdot q = 0\ , \quad q^2 =0\ , \quad V\cdot q =0 \ .
\end{equation}
Notice that this implies that $q$ is complex. These shifts do not alter
the kinematic invariants $s_{ij}$ and $t_{11}$ and  yield a linear $z$ dependence
in $t_{1i}, i=2,\ldots, m$. This is actually the minimal scenario where
total  (i.e.~bulk)  momentum is not conserved. We will call general
such shifts momentum non-conserving.

\paragraph{Two momenta shift.} Going over to the case where two momenta are affected, we can consider a shift
of the type (\ref{BCFWshift1}) but with different
deformation vectors $q_1, q_2$
\begin{equation}
\hp_1(z)= p_1 + q_1 z\ , \qquad \hp_2(z)= p_2 + q_2 z\ .
\end{equation}
In this case we need to impose the constraints
\begin{equation}\label{BCFWtcon1}
p_1\cdot q_1=p_2 \cdot q_2=0\ , \quad
\quad q_1^2=q_2^2=0
\end{equation}
as well as
\be V\cdot q_1 +V\cdot q_2=0.
 \ee
 If  $q_1+q_2 \neq 0$ we have an instance of momentum non-conserving shifts.
 Since the minimal such scenario is accommodated by a single shift, we will simplify
 the situation and restrict the two particle shifts to the momentum conserving
 case $q_1=-q_2=q$. Notice that this is the minimal scenario for momentum conserving
 shifts\footnote{In terms of our
pomeron analysis the minimal shifts, momentum conserving or not,
correspond to just two operators approaching each other which, in
the symmetric picture of \cite{Brower:2006ea}, means that the
scattering states split into two sets, one with only two operators
and another one with the rest.  For closed strings on the sphere
this holds independently for the holomorphic and anti-holomorphic
sector, while on the disc the holomorphic and anti-holomorphic
pieces are to be considered as two distinct operators on the
disc.} and it leaves unchanged the kinematic variables
$t_{ij},i,j =1,2$. The variables $s_{ik},t_{ik}$ with $i=1,2$ and
$k=3,\ldots,m$ are shifted linearly in $z$.

We still have the choice of imposing either $V \cdot q =0$ or $N\cdot q=0$.
If we do impose either one of these conditions then the variables
 $s_{ij}, i,j=1,  2$ remain unshifted. We will call such shifts {\em standard shifts}
 since they resemble the usual field-theoretic BCFW shifts.
Otherwise the $s_{ij}$ for $i,j=1,2$ are shifted quadratically in
$z$, which is not desirable if we wish the BCFW deformed amplitude
to have simple poles. Higher order poles lead to residues which
are non-rational functions of the kinematic variables i.e.~have
branch-cuts. Since this is not expected for tree level amplitudes
they should cancel among the various residues. However, this can
make the BCFW procedure rather cumbersome to implement. To avoid
this complication we can impose $(V\cdot q)^2=0$ (or,
equivalently, $(N \cdot q)^2=0$) so that the  $s_{ij},i,j=1,2$
invariants are shifted only linearly in $z$. Such shifts we will
call {\em s-shifts}.

Although in the previous discussion we considered scattering only
of bulk fields (closed strings), we
can obviously make similar two particle shifts for world-volume fields (open strings).
For the latter  we have the additional constraint that momenta are restricted on the
brane and the polarizations can be either parallel or transverse
to the brane. Furthermore, we can envisage more general shifts involving the momenta of
one bulk and one world-volume field but we will not delve into this.

In the presence of a $p-$brane the
spacetime Lorentz symmetry group  $SO(1,d-1)$ is broken to $SO(1,d-1) \rightarrow SO(1,p-1)
\times SO(d-p-1)$.
We can use the surviving symmetry
to bring the two particles to be deformed
to the following frame
\begin{equation}\label{BCFWmomt}
p_1 =(1, \lambda_1, 0, \dots; \lambda_{p+1},  \dots, \lambda_{d-1})\ , \quad
p_2 =(1, \mu_1, 0, \dots; \mu_{p+1},  \dots, \mu_{d-1}).\end{equation}
which in the massless case satisfy  $-1 + \sum_{i} \lambda_i^2 =-1+ \sum_i
\mu_i^2=0$. We have separated tangent and normal directions
by a semicolon. The deformation vector $q$
takes the general form
\begin{equation} \label{qt} q= (n_0, n_1 ,\ldots,n_p;x_{p+1},
\ldots,x_{d-1}).\end{equation} We see that  the shift vector  can have
components along the world-volume momenta unlike the usual
BCFW where a Lorentz transformation can bring them in
the form (\ref{mom}). This is not possible here due to the
presence of the defect which breaks the bulk Lorentz symmetry. Now
we discuss in more detail the various possibilities.

\subsubsection{One particle shift}

In this case we need to find a solution of the conditions
(\ref{oneparshi}). For a particle with momentum $p_1$ as in
(\ref{BCFWmomt}) we can make a shift, assuming without loss of
generality that $x_{p+3}=\dots=x_d=0$,
\begin{equation}\label{tshift}
q=(0,0,\dots; i{\lambda_{p+2} x_{p+3} \over \sqrt{\lambda_{p+1}^2
+\lambda_{p+2}^2}},-i{\lambda_{p+1} x_{p+3} \over \sqrt{\lambda_{p+1}^2
+\lambda_{p+2}^2}}, x_{p+3},0,\dots)\ .
\end{equation}
For the polarization vectors
\begin{equation}\label{the}
\he_1^-=q\ , \quad \he_1^+= q^* + z D\cdot p_1
 \end{equation}
the transversality
constraint $\hp_1 \cdot \he_1^+ =0$ demands
\begin{equation}\label{ttrans}
q\cdot q^*= 2 x_{p+3}^2=
-2p_1\cdot V\cdot p_1\ .
\end{equation}

\subsubsection{Two particle shifts}

As we discussed we distinguish two types of shifts where the  momenta of two particles
are deformed.

{\bf Standard shift.} This shift is identical to the shift in
(\ref{BCFWshift1}) and changes only $s_{ik}, t_{ik}$ for $i=1,2$
and $k=3,\ldots,m$. We can choose either  $x_{p+1}=\dots=x_d=0$ or
$n_0=n_1=\dots=n_p=0$. In the first case to satisfy the additional
constraints $p_1 \cdot q=p_2 \cdot q=0$ and $q^2=0$ we can select
$n_0=n_1=n_4=\dots=n_p=0$ and $n_2=i n_3$. We can find similar
solutions, albeit more complicated, in the second case provided
that $d-p \geq 3$.

{\bf S-shift.}  In this case, in addition to
$q \cdot p_1=q\cdot p_2 = 0$  and  $q^2 = 0$ we need to  impose
\begin{equation}\label{s2con}
(V \cdot q )^2=0,\quad {\rm with} \quad
p_i \cdot V\cdot q \neq 0\ , i=1,2 \qquad
\end{equation}
so that we obtain linear $z$ dependence in $s_{ij}, i,j=1,2$. A
solution of the above conditions is
\begin{equation}\label{s2shift}
q= (n_0, n_1, n_2, 0, \dots ; x_{p+1}, i x_{p+1}, 0,\dots)
\end{equation}
with
\begin{equation}
 n_0= \sqrt{n_1^2 +n_2^2}= { n_1 \big(\mu_1 (\lambda_{p+1} + i
\lambda_{p+2}) -\lambda_1 (\mu_{p+1} + i \mu_{p+2}) \big)
\over (\lambda_{p+1} -\mu_{p+1}) +
i (\lambda_{p+2} -\mu_{p+2})}\ , \quad x_{p+1}={ n_0-n_1 \lambda_1 \over
\lambda_{p+1} + i \lambda_{p+2}}\ .
\end{equation}
Transversality $\hp_1 \cdot \he_1^+ =0$ of the deformed particle's polarization
(\ref{gpolshift}) implies
\begin{equation}\label{s2trans}
 q\cdot q^*= -|n_0|^2
+|n_1|^2 + |n_2|^2= - p_1 \cdot p_2 = {t_{12} \over 2}
\end{equation} which allows us to determine $n_2$ (for example) and leaves
only one undetermined parameter.

For the $1\to 1$ scattering process of bulk fields off the brane
the above setup cannot be used. This is because we cannot bring
the two external particles back to back with equal energy as
before. In bulk scattering of course such an amplitude would be
just the propagator and has no meaning to make a BCFW shift. But
in the presence of branes we can indeed have such processes. In
this case we can use the symmetries of tangent and normal
directions  to bring the two momenta in the form
\begin{equation}\label{s3mom}
p_1=(1, \lambda_1, 0, \dots; \lambda_{p+1}, \cdots, \lambda_{d-1})\ , \quad
p_2=(-1, -\lambda_1, 0, \dots; \mu_{p+1}, \cdots, \mu_{d-1})\ .
\end{equation}
The two vectors satisfy momentum conservation along the brane but have
arbitrary normal components.

It turns out, see also subsection \ref{2gr}, that  for the level one states
$1\to 1$ scattering, the s-shift, which shifts linearly
the kinematic invariants, does not change the amplitude for all
possible polarization setups. So, in this setup it is more
convenient to relax the first condition in (\ref{s2con}). This
means that we will have second order poles when we use the BCFW
procedure. Nevertheless, as we mentioned before, the branch-cuts
in the residues will cancel leaving a rational function of
the external momenta.

A solution
of the conditions for this case is
\begin{equation}\label{s3shift}
q= (n_0, n_1, 0, \dots ; x_{p+1}, x_{p+2},0, \dots)
\end{equation}
with
\begin{eqnarray}
\quad n_0^2-n_1^2 &=& -\sigma^2\qquad \sigma \in \mathbb{R}\  ,
 \nonumber \\
n_0-n_1 \lambda_1 &=& { i \sigma (\lambda_{p+1}
\mu_{p+2}-\lambda_{p+2} \mu_{p+1})
\over \sqrt{(\lambda_{p+1} +\mu_{p+1})^2 +  (\lambda_{p+2} +\mu_{p+2})^2}}\ ,  \nonumber \\
x_{p+1}&=&{ i \sigma (\lambda_{p+2} + \mu_{p+2}) \over \sqrt{(\lambda_{p+1}
+\mu_{p+1})^2 + (\lambda_{p+2} +\mu_{p+2})^2}}\ , \nonumber \\
x_{p+2}&=&-{ i \sigma
(\lambda_{p+1} + \mu_{p+1}) \over \sqrt{(\lambda_{p+1} +\mu_{p+1})^2 + (\lambda_{p+2}
+\mu_{p+2})^2}} \ ,
\label{s3shiftrel}
\end{eqnarray}
and the additional constraint
from transversality
\begin{equation}\label{s3trans}
q\cdot q^*= -|n_0|^2
+|n_1|^2 + \sigma^2= - p_1 \cdot p_2 = {t_{12} \over 2}\ .
\ee
This last relation will be needed in the next section when we will
compare the large $z$ behavior of string amplitudes with the pomeron
results. Note also that in this case $q \cdot D \cdot p_1= - q
\cdot D \cdot p_2$.

\section{Pomerons and large $z$ behavior of open and closed (super) string amplitudes
}\label{Pom}

In this section we will employ the pomeron operators of section 3
to deduce the large $z$ behavior of  several string
amplitudes. Our analysis for
closed strings is very much parallel to the open string case
through the use of KLT relations \cite{Kawai:1985xq}. Nevertheless,
there are subtleties which appear and we comment upon them. For
open strings the situation is identical to the cases studied in
\cite{Cheung:2010vn,Boels:2010bv} and we will only cite their results.
We conclude this section with an explicit example where we
compute the BCFW behavior of the superstring scattering amplitude on the
disc of two level one closed string states \cite{Garousi:1996ad, Hashimoto:1996bf, Gubser:1996wt, klebanov} for
both the two particle and one particle shifts.

\subsection{Bosonic string amplitudes}\label{bosPom}

\paragraph{Open strings, two particle shifts.}

In this case the relevant pomeron operator  is (\ref{opbospom0})
and we shift the momenta so that  $k=\hk_1 +\hk_2= k_1+k_2$ and
$2Q= \hk_1-\hk_2= k_1-k_2 + 2q z$. The leading $ z$ dependence is
extracted when we replace $Q \to q z$, which implies that $|q| \gg
|k_1-k_2|$. Subleading terms should correspond to $\apr$
corrections as one can see in the supersymmetric pomeron
computation of \cite{Boels:2010bv}. When we insert the pomeron
operator, the  string amplitude takes the form
\begin{equation}\label{opbosM}
M^{\mu \nu}= z^{-2\apr k_1\cdot k_2} \left[ c z(-\eta^{\mu
\nu} + 2\apr k^\mu k^\nu)   + b (v^\mu
k^\nu - v^\nu k^\mu) + {\cal O}\left(\frac{1}{z}\right) \right]
\end{equation} where $v^\mu,
b, c$ are general functions depending only on  the
unshifted external momenta and polarizations.

To derive the large $z$ behavior of the S-matrix elements we need to
contract the expression above with the polarizations tensors
(\ref{gpolshift}) and utilize the Ward identities
\begin{equation}\label{ward} \hk_\mu^1 M^{\mu \nu} \he_\nu^2=0\  \Longrightarrow
\ q_\mu M^{\mu \nu} \he_\nu^2= - {1\over z} k^1_\mu  M^{\mu \nu}
\he_\nu^2
\end{equation}
as well as  the transversality identities $q\cdot k = q^*\cdot k=0$.
Actually it is the dependence of the antisymmetric part of
(\ref{opbosM}) on $k^\mu$ which results into different behavior
compared to field theory expectations for some polarization
configurations. The result appears in table 1
\cite{Boels:2010bv} for the behavior of bosonic string amplitudes
under BCFW shifts of level one open string states.
The polarizations $T'$ are orthogonal to $T$ and can be either
along the brane (gauge bosons) or normal to the brane (transverse
scalars).

\begin{table}\label{opbostable}
\begin{center}
\begin{tabular}{c|c c c}
$\e_1 \;\backslash \;\e_2 $ &  $-$  &  $+$  & T \\
\hline
$-$                           & $ z$ & $ {1\over z}$ & ${1\over z}$ \\
$+$                           & $ z^3$ & $z$ & $z$ \\
$T$                             & $z$ & $ {1\over z}$ & $z$ \\
$T'$                           & $z$ & $ {1\over z}$ & $   {1\over z}$ \\
\end{tabular}
\caption{The leading power in $z$ (modulo the overall factor
$z^{-\apr(k_1+k_2)^2}$) for adjacent
shifts of the all gluon (or transverse scalar) amplitude in the
{\em bosonic string} for all possible polarizations. The polarizations $T$ and $T'$ are orthogonal.}
\end{center}
\end{table}

\paragraph{Closed strings, two particle shifts.}
The operator in (\ref{bospomcl}) can be used to determine the
large $z$ behavior of string amplitudes on the sphere as in
\cite{Cheung:2010vn,Boels:2010bv} but also, with similar reasoning,
on the disc. On the disc, when the pomeron operator
(\ref{bospomcl}) is present, there are self-contractions as well
as contractions with other operators of the string amplitude. One
should make the substitutions (\ref{substitute}) in order to
compute the correlators.  For any case
the contractions of the pomeron with the other operators in the amplitude will
lead to terms like $ \apr Q\cdot p_k  \sim z$ and $
\apr Q\cdot D \cdot p_k \sim z$, where $k$ runs over all the
external particle momenta other than $k=1,2$.
Therefore a term in the pomeron proportional to $Q^x (D Q)^y$
will lead to an amplitude dependence as $z^{x+y}$.
There are
self-contractions as well which lead to terms such as $\apr Q\cdot D \cdot
Q\sim z$ for all shifts except the one we consider for $1\to 1$ scattering.
Hence, self-contractions in these amplitudes will lead to
a $z$-dependence of the form $z^u$ with $u < x+y$.
Finally, for the $1\to 1$ case there are only self-contractions
and the shift we consider leads to $\apr Q\cdot D \cdot
Q\sim z^2$. Moreover, in this particular case $Q\cdot D \cdot p=0$.

Based on the above, the final behavior of the amplitude is that dictated by KLT
\begin{eqnarray}\label{clbosM}
M^{\mu \bmu ;\nu \bnu} \sim z^{-\apr p_1\cdot p_2} \Big[ c z^2
P^{\mu \nu} \bar{P}^{\bmu \bnu} &+& b z (P^{\mu \nu} \bar{A}^{\bmu
\bnu}
+ A^{\mu \nu} \bar{P}^{\bmu \bnu} ) \nonumber \\ &+& (A^{\mu \nu; \bmu
\bnu}+ P^{\mu \nu} \bar{B}^{\bmu \bnu} + \bar{P}^{\bmu \bnu}
B^{\mu \nu}) +  {\cal O}\left(\frac{1}{z}\right) \Big]
\end{eqnarray}
where  $A^{\mu \nu}$ is an antisymmetric matrix as in
(\ref{opbosM}). We can write down the form of the $A_{\mu \nu}$ as
a sum of terms which originate from cross-contractions $A_1$ or
self-contractions $A_2$  of the pomeron vertex
\begin{eqnarray}\label{Amn}
A_1^{\mu \nu} \sim(v^\mu p^\nu - v^\nu p^\mu)\ , \quad
A_2^{\mu \nu} \sim (q^\mu p^\nu - q^\nu p^\mu)\ .
\end{eqnarray}
The matrix $A_2^{\mu \nu}$ comes from self-contraction of
$Q\partial X$ with $\bpartial X^\mu$.  The term $B^{\mu \nu}$ can
be written as the sum of two terms depending on the operator
contraction and has no symmetries.
The term $A^{\mu \nu; \bmu
 \bnu}$ comes from the contractions of $K^{\mu \nu} \bar{K}^{\bmu
\bnu}$. It is obviously antisymmetric in the  barred and
unbarred indices. In this case the self-contractions give
\begin{equation}
A_2^{\mu \nu; \bmu \bnu} \sim \big(D^{\mu \bmu} p^\nu (D\cdot
p)^{\bnu} +D^{\nu \bnu} p^\mu (D\cdot p)^{\bmu}\big) - \big(D^{\mu \bnu}
p^\nu (D\cdot p)^{\bmu}+ D^{\nu \bmu} p^\mu (D\cdot p)^{\bnu}\big)+
\dots
\end{equation}
which obviously has the same symmetries as implied by KLT
\footnote{We notice that self-contractions of pomeron operators
can lead to terms which contain the matrix $D^{\mu \bnu}$, therefore
breaking the product of the two independent left and right Lorentz
spin symmetries to a diagonal subgroup $SO(1,p)\times
\widetilde{SO}(1,p) \times SO(d-p-1)\times \widetilde{SO}(d-p-1)$.
Nevertheless these terms affect only the subleading in $z$
expression of the amplitude which, in any case, has the
aforementioned symmetries broken. Moreover these terms have the
same  symmetries in $(\mu, \bmu)$ and $(\nu, \bnu)$ as in the case
of supergravity in \cite{ArkaniHamed:2008yf} where no D-branes are
present.}.

\begin{table}
\begin{center}
\begin{tabular}{c c c c c c c }
\hline
$\epsilon_1 \backslash \epsilon_2$&$- -$&$- +$&$++$&$-$T&$+$T&TT\\
\hline \hline
$- -$&$z^2$&$1$&${1\over z^2}$&$1$&${1\over z^2}$&${1\over z^2}$\\
$- +$&$z^4$&$z^2$&$1$&$z^2$&$1$&$1$\\
$+ +$&$z^6$&$z^4$&$z^2$&$z^4$&$z^2$&$z^2$\\
$-$T&$z^2$&$1$&${1\over z^2}$&$z^2$ or $ 1$&$1$ or ${1\over z^2}$&$1$ or ${ 1\over z^2}$\\
$+$T&$z^4$&$z^2$&$1$&$z^4$ or $ z^2$&$z^2$ or $ 1$&$z^2$ or $ 1$\\
TT&$z^2$&$1$&${1\over z^2}$&$z^2$ or $ 1$&$1$ or ${1\over z^2}$&$z^2$, $1$, or ${1\over z^2}$\\
\end{tabular}
\caption{ The leading power in $ z$ (modulo the overall factor
$z^{-\hapr (p_1+p_2)^2}$) for the large $z$ limit of a an
amplitude under a graviton s-shift in the {\em bosonic string}
for all possible polarizations.}
\end{center}
\end{table}

\paragraph{Closed strings, one particle shift.}

This case is the single graviton shift and the relevant pomeron
operator is (\ref{bospomtad}). The large $z$ behavior is given by
the expression
\begin{equation}\label{tadbosM}
M_{\mu\bnu} = z^{-\hapr p \cdot D\cdot p} \left[ c z\big( -D_{\mu \bnu}
+ 2 \apr (V\cdot p)_\mu (V\cdot p)_{\bnu}\big ) + A_{\mu
\bnu} +  {\cal O}\left(\frac{1}{z}\right) \right]
\end{equation} where $A^{\mu \bnu} \sim v^\mu (V\cdot p)^{\bnu} -  (V\cdot p)^\mu
(D\cdot v)^{\bnu} $ is a generic matrix\footnote{Nevertheless by pulling a
$D_{\mu \bnu}$ matrix out of $M^{\mu \bnu}$ we can make it
antisymmetric.}
 whose explicit
dependence on $V\cdot p^\mu$ plays some role in the behavior of
amplitudes with $\e^T$ polarizations.

At this point we need to be careful with
the polarization assignment since the D-brane inverts the helicity
of the right-moving part of the operator. From the gluing of open
string operators we expect for the q-light cone polarizations
\begin{eqnarray}
\label{polass1}
(\he D)^{++}&=&
(q^* +zD\cdot p)(q^* -z p)\ , \nonumber \\
(\he D)^{-+}&=& q(q^* -zp)\ , \nonumber \\
 (\he D)^{+-}&=& (q^* +zD\cdot p)q\ , \nonumber \\
 (\he D)^{--}&=& qq\ .
\end{eqnarray}
which implies
\begin{eqnarray}
\label{polass2}
 \he^{++}&=& (q^*
+zD\cdot p)(q^* +z D\cdot p) \ , \nonumber \\
\he ^{-+}&=& q(q^* +zD\cdot p) \ , \nonumber \\
\he^{+-}&=& (q^* +zD\cdot p)q \ , \nonumber \\
\he^{--}&=& qq
\end{eqnarray}
where we have used the fact that the one-particle shift is  exclusively in the
normal directions (see (\ref{tshift})) and therefore $Dq=-q$. The
expression above agrees with the
transversality condition (\ref{ttrans}). Using Ward identities for gravitons analogous to
(\ref{ward}) we can deduce the large
$z$ behavior of table \ref{tadbostable}.

\begin{table}
\begin{center}
\begin{tabular}{c| c c c}
$\e_{\mu \bnu} $ &  $+$  &  $-$  & T \\
\hline
$-$                           & $ z$ & $ {1\over z}$ & ${1\over z} $ \\
$+$                           & $ z^3$ & $z$ & $z$ \\
T                             & $z$ & $ {1\over z} $ & $z$ \\
T'                            & $z$ & $ {1\over z} $ & $ { 1\over z }$ \\
\end{tabular}\caption{ The leading power in $z$ (modulo the overall factor
$z^{-\hapr p\cdot Dp}$) for the large $z$ limit of a an amplitude
under a graviton one-particle shift in the {\em bosonic string} for all
possible polarizations.}
\label{tadbostable}
\end{center}
\end{table}

\subsection{Superstring amplitudes}\label{susyPom}

\paragraph{Open strings, two particle shifts.}
To determine the large $z$ behavior in this case we use the pomeron
vertex operator of (\ref{opsusypom0}). It is straightforward to
derive
\begin{equation} \label{opsusyM} M^{\mu \nu} = z^{-2\apr  k_1 \cdot
k_2}\left[\eta^{\mu \nu}  \left(c z + c'  +{\cal O}\left(\frac{1}{z}\right) \right) +
A^{\mu \nu} + {B^{\mu \nu}\over z}+ \dots \right] \ ,
\end{equation} where $A^{\mu \nu}$ a generic antisymmetric matrix and $B^{\mu \nu}$ a general matrix\footnote{Notice that our symmetric OPE expansion does not yield in the pomeron vertex (\ref{opsusypom0})
the term with $\ddot X$ or that with $\dot\phi$,   as compared to (3.5) and (3.6) of
 \cite{Cheung:2010vn}.
However, we still get a contribution
of order $1$ multiplying $\eta^{\mu \nu}$, denoted by $c'$ in the previous formula,
from  $\eta^{\mu \nu} (Q\cdot \partial X)$ of  (\ref{opsusypom0})
since $Q=k_1-k_2+2qz$.
  Therefore,   we arrive at the same result as  that obtained
  with the pomeron vertex of  \cite{Cheung:2010vn} which was
  computed with a non-symmetric OPE.}.
Notice that it has similar structure to the bosonic case
but it differs in two crucial points: a) the leading term is
solely proportional to $\eta^{\mu \nu}$ and b) the subleading
antisymmetric matrix is generic and not of the special $k$-dependent form as in equation (\ref{opbosM}). These crucial
differences are responsible for the different behavior for some
polarization shifts compared to the bosonic case.
In table \ref{opsusytable} we present the large $z$ behavior for this case.

\begin{table}
\begin{center}
\begin{tabular}{c|c c c}
$\e_1 \;\backslash \;\e_2 $ &  $-$  &  $+$  & T \\
\hline
$-$                           & ${ 1\over z} $ & $ {1\over z} $ & ${1\over z} $ \\
$+$                           & $ z^3$ & $ {1\over z} $ & $ z$ \\
T                             & $ z$ & $ {1\over z} $ & $ z$ \\
T'                           & $ z$ & $ {1\over z} $ & $  1$ \\
\end{tabular}
\caption{The leading power in $z$ (modulo the overall factor
$z^{-\alpha'(k_1+k_2)^2}$) for the large $z$ limit of the adjacent shift
of an all gluon (transverse scalar) amplitude in the
{\em superstring} for all possible
polarizations.}\label{opsusytable}
\end{center}
\end{table}

\paragraph{Closed strings, two particle shifts.}
For this case we use the pomeron operator (\ref{susypomcl0}).
After contractions with the remaining operators in the path
integral or self-contractions in the $1\to1$ case we arrive at the
expected result (that is the product $M_{grav} \sim M_{gauge}
\times M_{gauge}$, where $M_{gauge}$ is given in (\ref{opsusyM}))
\begin{eqnarray}\label{clsusyM}
M^{\mu \bmu ;\nu\bnu}&\sim& z^{-\apr p_1\cdot p_2} \Bigg[ \eta^{\mu
\nu} \eta^{\bmu \bnu} z^2\left(c+ {\cal O}\left(\frac{1}{z}\right)\right) + b z (\eta^{\mu \nu}
\bar{A}^{\bmu \bnu} \nonumber \\
&+& A^{\mu \nu} \eta^{\bmu \bnu} ) + (A^{\mu
\nu ;\bmu\bnu} + \eta^{\mu\nu} \tilde{B}^{\bmu\bnu}+
\eta^{\bmu\bnu} B^{\mu\nu}) + {C^{\mu\nu;\bmu\bnu} \over z} +\dots
\Bigg]
\end{eqnarray}
with the matrices $A^{\mu \nu}\sim (v^\mu l^\nu -v^\nu
l^\mu)$ being generic antisymmetric matrices. This is in contrast to
the bosonic case where they depend on the pomeron momentum
$k$. Moreover cross- or self-contractions do not change the expected
symmetries,  from KLT relations, of these matrices. They have the
general structure
$$A^{\mu\nu ;\bmu\bnu} \sim A^{\mu \nu} \ \bar{A}^{\bmu
\bnu}\ ,$$
$$ C^{\mu \nu;\bmu \bnu} \sim \eta^{\mu \nu} (v^{\bmu} l^{\bnu}  -
v^{\bnu} l^{\bmu}) + [ (\mu ,\nu)  \longleftrightarrow  (\bmu,
\bnu)]\ .$$  $A^{\mu\nu ;\bmu\bnu}$ is antisymmetric in both $(\mu
,\nu)$ and $(\bmu, \bnu)$ while $C^{\mu\nu;\bmu\bnu}$ is a sum of
antisymmetric terms in each set. Notice also that as we mentioned
in the bosonic case the barred matrices are actually given in
terms of unbarred matrices i.e $\bar{G}^{\bmu \bnu}= D^{\bmu}_\a
D^{\bnu}_\b F^{\a \b}$. In any case the general symmetry
properties remain the same and an analysis along the lines of
\cite{ArkaniHamed:2008yf} leads to table \ref{clsusytable}.

\begin{table}
\begin{center}
\begin{tabular}{c|c c c c c c }

$\epsilon_1 \backslash \epsilon_2$&$- -$&$- +$&$++$&$-$T&$+$T&TT\\
\hline
$- -$&${1\over z^2}$&${1\over z^2}$&${1\over z^2}$&${1\over z^2}$&${1\over z^2}$&${1\over z^2}$\\
$- +$&$z^2$&$z^2$&${1\over z^2}$&$z^2$&$1$&$1$\\
$+ +$&$z^6$&$z^2$&${1\over z^2}$&$z^4$&$1$&$z^2$\\
$-$T&$1$&$1$&${1\over z^2}$&$1$&$1$ or ${1\over z}$&$1$ or ${1\over z}$\\
$+$T&$z^4$&$z^2$&$1$&$z^4$ or $z^3$&$1$&$z^2$ or $z$\\
TT&$z^2$&$1$&${1\over z^2}$&$z^2$ or $z$&$1$ or ${1\over z}$&$z^2$, $z$, or $1$\\
\end{tabular}\caption{The leading power in z (modulo the overall factor
$z^{-\hapr (p_1+p_2)^2}$) for the large $z$ limit of a an
amplitude under a graviton s-shift in the {\em superstring} for
all possible polarizations.}
\label{clsusytable}
\end{center}
\end{table}

\paragraph{Closed strings, one particle shift.}
 We insert (\ref{susypomtad}) and after we rename indices to make manifest the left-right spin
Lorenz symmetry, the leading $z$ behavior of the amplitude  is
\begin{equation}\label{tadsusyM}
M_{\mu \bnu} = z^{-\frac{\alpha'}{2}p \cdot D\cdot p} \left[  D_{\mu \bnu}\big(c z +
{\cal O}(1)\big) + A_{\mu \bnu} + {\cal O} \left( \frac{1}{z} \right) \right]
\end{equation} where $A^{\mu \bnu}$ is  an antisymmetric matrix in contrast to the similar matrix in the bosonic
which has an explicit dependence on the momenta. The large $z$
behavior of amplitudes under this shift is given in table
\ref{tadsusytable}.

\begin{table}
\begin{center}
\begin{tabular}{c| c c c}
$\e_{\mu \nu} $ &  $+$  &  $-$  & T \\
\hline
$-$                           & $ {1\over z}$ & $ {1\over z} $ & ${1\over z} $ \\
$+$                           & $ z^3$ & ${1\over z}$ & $z$ \\
T                             & $z$ & $ {1\over z} $ & $z$ \\
T2                            & $z$ & $ {1\over z} $ & $  1 $ \\
\end{tabular}\caption{ The leading power in z (modulo the overall factor
$z^{-\hapr p\cdot Dp}$) for the large $z$ limit of a an amplitude
under a graviton t-shift in the {\em superstring} for all
possible polarizations.}
\label{tadsusytable}
\end{center}
\end{table}

\subsection{An  example: graviton scattering off a D-brane}\label{2gr} In
this subsection we will work explicitly on the two graviton
superstring scattering off a D-brane to verify the behavior
advocated in the previous sections under the BCFW shifts. We begin
by writing down the amplitude 
 \cite{Garousi:1996ad, Hashimoto:1996bf, Gubser:1996wt, klebanov}
\begin{equation}\label{amp2gr}
{\cal A}(p_1,p_2) \sim -i\frac{ \kappa\,T_p}{2}\,\frac{\G\left(-{\apr \over 4} t\right)\G\left(\hapr
s\right)}{\G\left(1-{\apr \over 4} t+\hapr s\right)}
\left(s\,a_1+\frac{t}{2}\,a_2\right)
\end{equation}
where $t=-(p_1+p_2)^2=-2p_1 \cdot  p_2$ is the momentum transfer
to the $p$-brane and $s=p_1\cdot D\cdot p_1=2p_1\cdot V\cdot p_1$
is the momentum flowing parallel to the world-volume of the brane.
The kinematic factors above are:
\begin{eqnarray}\label{defa}
a_1&=&{\rm tr}(\pol_1\cdot D)\,p_1\cdot \pol_2 \cdot p_1
-p_1\cdot\pol_2\cdot D\cdot\pol_1\cdot p_2 -
p_1\cdot\pol_2\cdot\pol_1^T \cdot D\cdot p_1
\nonumber\\
&&\ -p_1\cdot\pol_2^T \cdot \pol_1 \cdot D \cdot p_1 -
p_1\cdot\pol_2\cdot\pol_1^T \cdot p_2 + {s\over 2}\,{\rm
tr}(\pol_1\cdot\pol_2^T) +\Big\{1\longleftrightarrow 2\Big\}\ ,
\\\nonumber\\
a_2&=&{\rm tr}(\pol_1\cdot D)\,(p_1\cdot\pol_2\cdot D\cdot p_2 +
p_2\cdot D\cdot\pol_2\cdot p_1 +p_2\cdot D\cdot\pol_2\cdot D\cdot
p_2)
\nonumber\\
&&+p_1\cdot D\cdot\pol_1\cdot D\cdot\pol_2\cdot D\cdot p_2
-p_2\cdot D\cdot\pol_2\cdot\pol_1^T\cdot D\cdot p_1 +{s\over
2}\,{\rm tr}(\pol_1\cdot D\cdot \pol_2\cdot D)
\nonumber\\
&&-{s\over 2}\,{\rm tr}(\pol_1\cdot\pol_2^T) -{\rm tr}(\pol_1\cdot
D) {\rm tr}(\pol_2\cdot D)\, \left({s\over 2}-\frac{t}{4}\right)
+\Big\{1\longleftrightarrow 2 \Big\}\ \ . \end{eqnarray} Our
notation is such that, for example,  $p_1\cdot\pol_2\cdot\pol_1^T \cdot
D\cdot p_1
=p_1^\mu\,\pol_{2\mu\nu}\,\pol_1{}^{\lambda\nu}\,D_{\lambda\rho}\,p_1^\rho$.

\subsubsection{Two particle shift: s-shift}

The standard BCFW shift is not applicable here since we have only
two scattered states. We have to study therefore the s-shift under which
we have
\begin{equation}\label{s3grmom}
\hs= s+ 4 (p_1\cdot V\cdot q) z + 2 (q\cdot V\cdot q) z^2\ , \quad
\hht= t\ .
\end{equation}
The gamma function prefactors have an expansion for large $a=
{2\sqrt{2} p_1\cdot V\cdot q \over \sqrt{q\cdot V\cdot q}}$ and
$\hz= z \sqrt{2  q\cdot V\cdot q}$ as
\begin{equation}\label{s3gamexp}
\frac{\G\left(-{\apr  \over 4}t \right)\G\left(\hapr \hs\right)}{\G\left(1-{\apr \over
4}t +\hapr \hs\right)} \sim \Gamma\left(-{\apr  \over 4}t\right)\hz^{\hapr t-2}\left[
1+ a \left({\apr  \over 4}t-1\right) {1\over \hz} + {\cal O}\left({1\over \hz^2}\right)\right]\ .
\end{equation}

Here we need to make a point. The shift is quadratic in $z$ so we
might wonder whether we will have branch cuts in the complex
momentum plane when we try to derive BCFW relations. Certainly for
tree level amplitudes this would be bizarre. The gamma functions
has two poles from the two solutions of $\hapr \hs=-n, \ n\in
\mathbb{N}$. The poles are given generically by an expression
which contains square roots of the kinematic variables and the two
solutions differ by the branch of the square root. At least at the
level of the gamma functions the only other part, except the
denominator, depends on $\hz$ is $\Gamma\left(1-{\apr  \over
4}t +\hapr \hs\right)\to \Gamma\left(1-{\apr  \over 4} t-n\right)$, that is
independent of the residue position $\hz$ itself. Therefore the
final result would be the sum of the two solutions which differ by
the branch of the square root and should produce a rational
function of the external momenta \cite{Cachazo:2005ga}. Definitely
this point needs further clarification but in lack of the BCFW
recursion relations themselves further understanding is difficult.

 We will study now  some of the possible polarization assignments.
\begin{itemize}
\item $\he_1^{--}=q q$ and $\he_2^{++}= qq$. It is easy to see
that $a_1$ in (\ref{amp2gr}) vanishes and only $a_2$ contributes.
Calling $K^{\mu \bmu;\nu \bnu}$ the kinematic factor of the
amplitude, we have
\begin{equation}\label{Kmmpp}
K^{--;++} \sim (q\cdot D\cdot q)^2 {t^2\over 4}
\end{equation}
where we have used the relation $q\cdot D\cdot p_1=-q\cdot D\cdot
p_2$. Upon combining with (\ref{s3gamexp}) we indeed get
\begin{equation}\label{Mmmpp}
M^{--;++} \sim \hz^{-\apr p_1\cdot p_2-2}\Big((q\cdot D \cdot q)^2
 {t^2\over 4} + \dots\Big)
\end{equation}
as expected from table \ref{clsusytable}.

 \item $\he_1^{--}=q q$ and $\he_2^{--}= (q^*-zp_1)(q^*-zp_1)$. In
 this case we use $q^* q = -p_1\cdot p_2$ and
 \begin{equation}
 q\cdot D\cdot q= q^* \cdot D \cdot q^*,  \quad  q^* \cdot D\cdot q = qq^*- q\cdot D\cdot
 q,
 \quad q^*\cdot D\cdot p=-q\cdot D\cdot p\ .
  \end{equation}
 After some algebra we get
\begin{equation}\label{Kmmmm}
K^{--;--} \sim {t^2\over 4} (q\cdot D\cdot q-s)^2  + \dots \ .
\end{equation}
Combined with (\ref{s3gamexp}) it gives the expected behavior for
the amplitude $M^{--;--}\sim z^{6}$.

\item $\he_1^{++}= (q^*+zp_2)(q^*+zp_2)$ and  $\he_2^{++}=q q$. This is
similar  to the previous case and leads to the same result.

\item $\he_1^{++}= (q^*+zp_2)(q^*+zp_2)$ and $\he_2^{--}=
(q^*-zp_1)(q^*-zp_1)$. After some straightforward calculations
we find
\begin{equation}\label{Kppmm}
K^{++;--} \sim \hz^8 (q\cdot D\cdot q)^2 \halft^4
\end{equation}
which agrees with expectations. The remaining cases can be
verified in a similar manner as well.
\end{itemize}

\subsubsection{Single particle shift}

In this case the shift of the momentum $p_1$ results in
\begin{equation}\label{tgrmom}
\hs= s\ ,  \qquad \hht= t -2 (q\cdot p_2) z= t-\hz
\end{equation}
where $\hz = 2  q\cdot p_2 z$. The prefactor expansion is
\begin{equation}\label{tgamexp}
\frac{\G\left(-{\apr  \over 4} \hht\right)\G\left(\hapr s\right)}{\G\left(1-{\apr \over 4} \hht+\hapr s\right)} \sim \Gamma\left(\hapr s \right)
\hz^{-\hapr s-1}\left[ 1- {\apr \over 2 \hz ^2 }\left(1+\hapr s\right)(s-t)
+ {\cal O}\left({1\over \hz^3}\right)\right]\ .
\end{equation}
We also need to choose a gauge for $\e_2^{\mu\nu}$. A convenient
choice is to choose a reference momentum vector $q$ and impose the conditions
\begin{equation}\label{qgauge}
q_\mu e_2^{\mu \nu}=0\ .
\end{equation}

\begin{itemize}
\item $\e_1^{--}= qq$. The result is given by
\begin{equation}\label{Kmm}
K^{--}\sim s\;  {\rm tr}(\pol_2\cdot D) (q\cdot p_2)^2
\end{equation}
which leads to the expected $\hz^{-\hapr s-1}$ behavior according
to table \ref{tadsusytable}.

\item $\e_1^{-+}= q(q^* + D\cdot p_1)$. After some relatively
lengthy calculation we get
\begin{eqnarray}\label{Kmp}
K^{-+}&\sim & s\; \Big[ 4 s (p_1\cdot \pol_2\cdot p_1) - 4 (p_2\cdot
q) q^* \cdot \pol_2 \cdot D\cdot p_2 + 2t (p_2\cdot D +p_1)\cdot
\pol_2 \cdot D\cdot p_2 + \nonumber \\  &&+{\rm tr}(\pol_2\cdot D)
\big(4(q\cdot p_2)(q^*\cdot p_2)+ t(-2s+t)\big) + \dots \Big]
\end{eqnarray}
which gives again $\hz^{-\hapr s-1}$ as required.

\item $\e_1^{++}= (q^* + D\cdot p_1)(q^* + D\cdot p_1)$. We find
\begin{equation}\label{Kpp}
K^{++}\sim z^3 s \; {\rm tr}(\pol_2\cdot D) (q\cdot p_2)^2
\end{equation}again in agreement. The remaining polarization choices
require a bit more work but are in agreement with the pomeron analysis
expectations.
\end{itemize}

\section{Field theory
versus pomeron in the eikonal Regge regime }\label{FT}

\subsection{General remarks}

In this section we will try to identify a field theory which leads
to amplitudes with the same large $z$ behavior under BCFW
deformations as the pomeron analysis in a special kinematic regime.
We distinguish our fields in  a
``soft" classical background, which describes the unshifted particles, and a
``hard"  one corresponding to  the BCFW shifted particles that carry
large momentum parametrized by $z$.  In the limit $z \rightarrow \infty$
we can consider the scattering amplitude as a process where 
the hard particle is shooting through the  soft background. The large $z$
behavior of this amplitude can be determined by analyzing  
the quadratic fluctuations around the soft background in an appropriate two
derivative Lagrangian.  

Let us consider the 
scattering of a number of open and closed strings with momenta $k_I$ and $p_J$ respectively. 
We now assume that $\sqrt {\alpha' } k_I, \
\sqrt{\alpha' } p_J \sim \mathcal{O}(\epsilon)$ with $\epsilon << 1$
for all $I, J$, while
  $\sqrt {\alpha' } q \sim
\mathcal{O}(\epsilon^{-1})$ so that $q \cdot k_I$ and $q\cdot p_J$ is of order 1.
Subsequently, we take $z$ to be large. Then  $\alpha' s_{IJ}
\ll 1$ while $\alpha' \hat s_{IJ} \sim z$ and the
pomeron approximation is valid. 
We will refer to this particular
limit of parameter space
 as the eikonal Regge (ER) regime
\cite{Cheung:2010vn}.
 Notice that this regime is clearly distinct from the
field theory one which corresponds to  $\apr \to 0$ and for which
all kinematic variables are small compared to the string
scale.

An obvious candidate field theory
is an appropriate truncated version of  the DBI
supplemented by the supergravity effective action\footnote{Notice that
we ignore all the higher derivative
corrections to the DBI action since they would definitely spoil BCFW constructibility.}.
In appendix C we give
a detailed description of this action. We stress the truncated
version of the DBI action because, as pointed out in
\cite{ArkaniHamed:2008yf}, higher dimension operators such as the
multipole type couplings present in the DBI action spoil the good
behavior of  SYM amplitudes under BCFW deformations. We
need to think therefore of a two derivative action which will be the
appropriate truncation of the DBI action.

Another important point is that when we consider tree level field
theory diagrams constructed by D-brane  and bulk fields, not
all of them correspond to tree level processes in the string
theory side. This is exactly what happens if we consider, for
instance, the exchange of a graviton between two branes.
It corresponds to an annulus amplitude in string theory and
therefore, due to world-sheet duality, to an 1-loop
open string amplitude. There is no clear separation of tree and
loop level processes in string theory as seen from the field
theory side. The only diagrams where bulk fields can be used as
intermediate states are those between a brane vertex and a bulk
one. Hence, we will need to restore Newton's constant $\kn$ in the supergravity
and brane actions in order to keep track of such diagrams.

 We also point out that in the following analysis the soft background fields
depend on the coupling constants. These  fields are
solutions of the corresponding equations of motion that
non-linearly complete the sum over the plane waves describing the
soft external particles \cite{ArkaniHamed:2008yf}. For example, a
background such as the metric  can be written as $g_{\mu
\nu}=\eta_{\mu \nu} + H_{\mu \nu}(p_i,\epsilon_i,\kn)$. Therefore,
a soft background field might actually vanish  if we take some coupling
constants to zero.

\subsection{Field theory analysis of open string BCFW
deformation and single bulk field shift} 
As we explained earlier, the goal is
to identify a two derivative action which
describes hard particles moving through a soft background
and leads to the same large $z$ amplitude behavior as the
pomeron analysis in the ER regime.
 We will assume that in the field theory setup
one needs to retain only the DBI terms quadratic in the hard
fields.  This analysis is presented in appendix C. The
soft background has an arbitrary number of soft fields which
describe the  unshifted particles of a given amplitude. So we
use as starting point the Lagrangian (\ref{qscalar}) or its
equivalent form for gauge fields which we have not written down.

\paragraph{Non-dynamical metric: pure brane field amplitudes.}
Let us consider first the case of a non-dynamic metric which
corresponds to amplitudes with only world-volume fields.
The action in (\ref{qscalar}) takes a very simple form
with the induced metric given by \begin{equation}\tg_{\a \b} =
\eta_{\a \b} +
\partial_\a x^i \partial_\b x_i . \end{equation}
 We see that the leading coupling in (\ref{qscalar}) is at least
quartic in the fields having the two hard fields $\lambda^i$ 
and at least two soft
ones $x^i$.  
Since we have already
fixed the gauge we cannot impose in addition 
the $q$-light cone gauge  and therefore  we
have a vertex at least quadratic in the soft fields which goes as
$\mathcal{O}(z^2)$. On the contrary,  the pomeron analysis in
(\ref{opsusyM}) in the ER regime gives a maximum $z$ dependence
$\mathcal{O}(z)$.



The problem above persists in the non-abelian case too.
 It is known
\cite{ArkaniHamed:2008gz, Boels:2010bv} that the pomeron analysis
for non-adjacent shifts in color ordered amplitudes produces
subleading large $z$ behavior compared to the adjacent case we
considered in this paper. So, the upper limit of the pomeron
analysis is still the $\mathcal{O}(z)$ behavior of (\ref{opsusyM})
in the ER regime. On the other hand, the DBI analysis for the
non-abelian theory will result in an $\mathcal{O}(z^2)$ behavior
since this is due to the kinetic term common to both abelian and
non-abelian actions.  In any case, the DBI theory includes 
higher dimension brane-field operators that lead to increasingly
divergent vertices for hard fields therefore spoiling BCFW
constructibility \cite{ArkaniHamed:2008gz}. Obviously this behavior is not compatible 
with the pomeron analysis.

\paragraph{Dynamical metric: mixed brane and bulk field amplitudes.}
Consider now the case where the metric is dynamic. In the
previous paragraph we concluded that brane field operators with 
dimension higher than quartic
are problematic due to the induced metric expansion
in the action.  One might wonder what will happen
if we consider amplitudes which involve the two hard
scalars/gauge bosons and $n$ soft gravitons. This case
could  work since we need only the bulk metric
without additional soft brane fields.

The analysis proceeds as in the photon-graviton case of 
\cite{ArkaniHamed:2008yf}. If we can choose the  $q$-light cone gauge 
the vertices of order ${\cal O}(z^2)$ can be eliminated using 
the gauge symmetry of the bulk graviton background.
Otherwise there is a unique set of diagrams where 
the two hard scalars are connected via a single cubic vertex
to the soft graviton background field. The latter is subsequently
 connected to the soft gravitons through  bulk interactions.
 In this case we obtain an  $\mathcal{O}(z^2)$  dependence
 which again disagrees with the pomeron result (\ref{opsusyM})
 in the ER regime.

 To track the origin of this discrepancy we can push this analysis
a bit further. Consider the amplitude of two scalars carrying momenta $k_1, k_2$ and one
graviton with momentum $p$. This is
given by the expression \cite{Hashimoto:1996kf} 
\begin{equation}\label{2sc1gr}
A\sim {\Gamma(-\apr t) \over \Gamma\left(1-\hapr t\right)  }t
\left (-4 \apr (\epsilon_1 \cdot \epsilon_2) \ (k_1^\mu
\epsilon_{\mu \nu} k_2^\nu) + \hapr (\epsilon_1\cdot
p)(\epsilon_2\cdot p) {\rm tr}(\epsilon D)+\dots \right)\ ,
\end{equation}
where we have provided only the relevant piece  of the corresponding
kinematic factor  for our discussion. As in the tachyon discussion
in subsection \ref{PomEx} there is no Regge behavior for
this amplitude\footnote{For the tachyon amplitude there is no
kinematic factor but only a gamma function prefactor which
is not shifted under this BCFW deformation and therefore the
amplitude does not have any $z$ dependence.}.   
Nevertheless, if we use the BCFW standard shift of subsection
\ref{BCFWdef} for open strings in the ER limit, from the first term of the
amplitude above we get an
 $\mathcal{O}(z^2)$ behavior in agreement with the aforementioned
 cubic vertex of DBI. We see that the two scalar/one graviton vertex,  which would lead
to the unique set of diagrams we mentioned above, cannot be
computed using the pomeron method. This leads in essence to the
discrepancy between field theory and pomeron analysis.

Moreover if we use the momentum non-conserving deformation of a
single graviton of subsection \ref{BCFWdef}, we obtain from the
second term in (\ref{2sc1gr}) an $\mathcal{O}(z^2)$ behavior. This
pattern for the single particle shifts can be easily
generalized to higher point functions. The amplitude of
three scalars and one graviton computed in
\cite{Fotopoulos:2001pt}, like the tachyon amplitude in
(\ref{3t1T}), has no Regge behavior under the single particle
shift. Nevertheless, terms such as $(\epsilon \cdot p)^3 {\rm
tr}(\epsilon D)$ in the kinematic factor lead to
$\mathcal{O}(z^3)$ behavior under these BCFW shifts, suggesting we
keep terms with more than two derivatives,  schematically of the form  $x^3\partial^3
h$, from Taylor expanding the metric in the DBI action as in
(\ref{Taylorg}). Proceeding further, the one graviton with $n$
scalars amplitude will have a $\mathcal{O}(z^n)$ behavior under
the one particle shift making the divergence at complex infinity
worse for higher point amplitudes. Therefore,
there are shifts of string
amplitudes  which do not vanish at complex
infinity even away from the ER regime, and therefore spoil constructibility 
under these BCFW deformations.

The final conclusion of this subsection is that
the ER regime conjecture is not valid  for
mixed open-closed string amplitudes without making further
assumptions. It might be however that an extra  refinement of the ER limit
can lead to agreement between a proper truncation of the DBI
theory and the pomeron analysis. For instance, the obvious way to achieve agreement is by considering 
the limit that eliminates all higher
dimension operators on the D-branes as well as their
coupling to supergravity modes.
As
we can easily see from (\ref{constants}), for D$p$-branes with $p\le
3$ we can take the limit $\apr \to 0$ and $g_s \to 0$ in order to
keep $g_{Dp}$ fixed but $\kn \to 0$. This is actually the usual
decoupling limit in AdS/CFT correspondence \cite{Aharony:1999ti} where the system
is reduced to two independent subsystems:
 a) an  abelian gauge theory along with some free
scalars on the D-brane and b) the bulk gravity theory.

Notice that in this decoupling limit  
we need to consider the non-abelian theory on
a stack of  D-branes in order to end up with  an interacting theory.
For gauge bosons the analysis is
identical to that of
\cite{ArkaniHamed:2008yf} since the DBI action yields the 
Yang--Mills Lagrangian to quadratic level. The subleading
antisymmetric terms in \cite{ArkaniHamed:2008yf} come from the
commutators in the gauge field strength. This agrees with the
subleading piece of the pomeron computation (\ref{opsusyM}).
For the case of transverse scalars similar antisymmetric
terms come from the commutators of the non-abelian DBI as in
\cite{Myers:1999ps}. The Lagrangian can be written schematically
\begin{equation}\label{Lopen}
L=  {\rm tr}\left( F_{\a \b} F^{\a \b} +   (D_{\a} X^i)^2 +   [X^i,
X^j]^2\right) \ ,
\end{equation} where $D_{\a} X^i= \partial_{\a} X^i + [A_{\a},X^i]$.
Based on the remarks above the analysis is very similar to the pure YM
case which has appeared in the literature and we will not repeat it
here. It leads to  table \ref{opsusytable} in agreement with
the pomeron results.

\subsection{Field theory analysis of closed string BCFW
deformations.} It is will be also instructive to try to understand the problem
at hand from the point of view of bulk shifts like those in
subsection \ref{PomEx}.
As in
\cite{ArkaniHamed:2008yf} we define vielbeins $e, \tilde{e}$, with
their associated connections $\omega, \tilde{\omega}$ and we
write 
\be 
h_{\mu \nu} = e^a_\mu \tilde{e}^{\tilde{a}}_\nu h_{a
\tilde{a}} , \ \nabla_\kappa h_{\mu \nu} = e^a_\mu
\tilde{e}^{\tilde{a}}_\nu D_\kappa h_{a \tilde{a}} \ee with \be
D_\kappa h_{a \tilde{a}} =
\partial_\kappa h_{a \tilde{a}} + {\omega^b_{\kappa a}} h_{b
\tilde{a}} + {\tilde{\omega}^{\tilde{b}}_{\kappa \tilde{a}}}h_{a
\tilde{b}} . 
\ee  
The action  becomes
\begin{eqnarray}\label{Lagsshift} S &=& \int d^{d}x  \sqrt{-g} \left( G_{E}^{a \tilde{b}} h_{a \tilde{b}}+ \frac{1}{4}
g^{\mu \nu} \eta^{ab} \tilde{\eta}^{\tilde{a} \tilde{b}} D_\mu
h_{a \tilde{a}} D_\nu h_{b \tilde{b}} - \frac{1}{2} h_{a
\tilde{a}} h_{b \tilde{b}} R^{a
b \tilde{a} \tilde{b}}\right) + \nonumber \\
&& {1 \over (\apr)^2 g^2_{Dp}}
\int d^{p+1}\sigma \ \left(\kn \tilde{h}_{\a \b} T_{DBI}^{\a \b} + \kn^2
\tilde{h}_{\a \b} \tilde{h}_{\gamma \delta}Z_{DBI}^{\a \b; \gamma
\delta}\right)  \ ,
\end{eqnarray} where $G_{E}^{a \tilde{b}}$ is the Einstein tensor
corresponding to the equations of motion of the background.
In the notation above we are trying to make evident the two
distinct Lorentz spin symmetries acting on the left and right
index of $h_{\mu\nu}$. For the pull-back tensors we use the
pull-back vielbeins $\tilde{e}^a_\alpha =
\partial_\alpha x^\mu e^a_\mu$. It should be clear  from our earlier
discussion that the supergravity action, while it naively looks
independent of $\kn$, it does have an implicit dependence through
the graviton background fields.

For the momentum conserving two particle shifts the field theory
analysis requires tadpole cancellation. The term linear in $h_{a
b}$ is a tadpole due to the presence of the boundary state meaning 
that the DBI action sources the Einstein equations
\begin{equation}\label{Ein}
G_{E}^{a \tilde{b}}= T_p \ \delta^{d-p-1}(x^i-x_0^i) \
\tilde{T}_{DBI}^{a \bar{b}}  \ .
\end{equation}
We assume that the deformation vector $q_\mu$ is along the $+$
direction and we choose the light cone gauge
 \be
 \omega^+_{a
b} = \tilde{\omega}^+_{\tilde{a} \tilde{b}} = g^{++} = g^{+
\kappa} = 0 \ \ \ \mathrm{and} \ \ \ g^{+-} = 1 ,
\ee
so that there
are no $O(z^2)$ vertices.

For the moment let us ignore the quadratic graviton contribution of
the DBI action. Then the analysis proceeds exactly as in
\cite{ArkaniHamed:2008yf}.  The order ${\cal O}(z)$
vertices preserve both the spin Lorentz symmetries -- except for
the unique set of bulk diagrams (not to be confused with the 
unique brane diagrams which play role for the DBI couplings)  that give
contributions up to ${\cal O}(z^2)$. These ${\cal O}(z^2)$ terms come from the
two derivative part of the Lagrangian, do not break either of the
left or right spin ``Lorentz" invariance, and  thus are
proportional to $\eta^{ab} \tilde{\eta}^{\tilde{a} \tilde{b}}$.
The order  ${\cal O}(z)$ terms that violate the symmetry and come from a
derivative on $h$ and a single $\omega$ or $\tilde{\omega}$
insertion have the form $\eta^{ab} \tilde{A}^{\tilde{a}
\tilde{b}} + A^{a b} \tilde{\eta}^{\tilde{a} \tilde{b}}$ where
$A$ and $\tilde{A}$ are antisymmetric.
The ${\cal O}(1)$ parts of the amplitude will
receive contributions from both the bulk and  the DBI
action. Most of the pieces and their origin is the same as in
\cite{ArkaniHamed:2008yf} and we will refer the reader to this
paper for further details.

The bulk Riemann tensor term of the action is sourced by the
stress energy tensor (\ref{Ein})   of the DBI action. The Einstein
equations give a backreacted metric and a corresponding
Riemann tensor. This will not affect the large $z$ behavior of the
amplitudes since they depend on the symmetries of the Riemann
tensor rather than its actual form. Nevertheless the $D^{a
\tilde{a}}$ dependence of the matrix elements of $M^{a \tilde{a} b
\tilde{b}}$   in the pomeron analysis (\ref{clsusyM}) has its
origin on the brane contribution to the background metric. The
brane with all its fields is part of the soft background through
which the hard gravitons propagate\footnote{Considering only
bosonic fields for simplicity we can see that the D-brane
contributions to (\ref{clsusyM}) are of subleading order, ${\cal
O}(z)$ and bellow, and have the same structure with the ${\cal
O}(z)$ terms of the pure gravity analysis. This is very similar to
equation (4.16) of \cite{Cheung:2008dn} which has given the large
$z$ analysis of gravity coupled to matter.}. Thus we find \be
\label{clFTM} M^{a \tilde{a} b \tilde{b}} = c z^2 \eta^{ab}
\tilde{\eta}^{\tilde{a} \tilde{b}} + z \left(\eta^{ab}
\tilde{A}^{\tilde{a} \tilde{b}} + A^{ab} \tilde{\eta}^{\tilde{a}
\tilde{b}} \right) + A^{a b \tilde{a} \tilde{b}} + \eta^{ab}
\tilde{B}^{\tilde{a} \tilde{b}} + B^{ab} \tilde{\eta}^{\tilde{a}
\tilde{b}} + \frac{1}{z} C^{a b \tilde{a} \tilde{b}} + \cdots \ee
where $A^{ab}$ is an antisymmetric matrix, $B^{ab}$ is an
arbitrary matrix, and $A^{a b \tilde{a} \tilde{b}}$ is
antisymmetric in $(ab)$ and $(\tilde{a} \tilde{b})$. This agrees
with the pomeron result (\ref{clsusyM}).
From the point of view of field theory it is quite
remarkable that this symmetry structure is precisely what we
get  by squaring the Yang-Mills result (\ref{opsusyM}).
Instead,  this phenomenon is perfectly understood in string theory
in light of the KLT relations \cite{Kawai:1985xq}.

Finally, notice that if we do not ignore the quadratic graviton
contribution of the DBI action we face an obvious problem. The
tensor $Z_{DBI}^{\a \b;\gamma \delta}$ has no symmetry in its
world-volume indices. This term gives a contribution to the
$\mathcal{O}(1)$ terms of (\ref{clFTM}) but it does not have the
same symmetries as any of them and so it will spoil agreement with the
pomeron result (\ref{clsusyM}).
Therefore, we need to consider the
decoupling limit once more. For $p<3$ this limit
requires $g_s \to 0$ as well, while  for $p=3$ we can keep $g_s$ fixed
while sending $\kn \to 0$ and keeping $g_{D3}$ fixed. Then one notices
that the tadpole term in (\ref{qgrav}) survives and the quadratic
one vanishes in accord with (\ref{clsusyM}).

\acknowledgments We would like to thank C.~Angelantonj, T.~Taylor
and M.~Tsulaia for valuable discussions. The work of A.~F. was
supported by an INFN postdoctoral fellowship and partly supported
by the Italian MIUR-PRIN contract 20075ATT78. The work
of N.~P was partially supported by the Swiss National Science
Foundation.

\renewcommand{\thesection}{A}
\setcounter{equation}{0}
\renewcommand{\theequation}{A.\arabic{equation}}
\section*{Appendix A: Details on pomeron derivations}\label{A}

In this appendix we give a detailed derivation of the bosonic
closed string  pomeron vertex for level 1 states. The
supersymmetric case follows in a similar manner. In the formulas
below we show only the leading terms in the shifted momenta.
Subleading terms in $\hp_i$ or $\hk_i$ are represented by
ellipsis. We begin with the OPE
\begin{eqnarray}\label{grOPE1}
&&
\partial X^\mu \bpartial X^\nu e^{i \hp_1 \cdot
X}\left({w\over 2}\right)
\partial X^\rho \bpartial X^\sigma e^{i
\hp_2 \cdot X}\left(-{w\over 2}\right)\sim
 e^{i p \cdot X + i w Q \cdot \partial X + i\bw Q\cdot \bpartial X }(0) |w|^{\apr p_1\cdot p_2}
 \nonumber \\&&
 \Big(
{-\eta^{\mu \rho} + \hapr \hp_2^\mu \hp_1^\rho \over w^2} + i {
\partial X^\mu \hp_1^\rho - \partial X^\rho \hp_2^\mu \over w} +
\dots \Big)
\times
 \Big( {-\eta^{\nu \sigma} +
\hapr \hp_2^\nu \hp_1^\sigma \over \bw^2} + i { \bpartial X^\nu
\hp_1^\sigma -
\partial X^\sigma \hp_2^\nu \over \bw} + \dots \Big)\nonumber \\
\end{eqnarray}
where, as in the main text, $p=\hp_1+\hp_2$ and $2Q= \hp_1-\hp_2$.
Now we make the substitutions
\begin{equation}\label{grOPE2}
W=- w Q\cdot \partial X\ , \qquad \bar{W}= -\bw Q\cdot \bpartial X
\end{equation}
and we integrate over the position $W= r e^{i\theta}$ on the sphere
using the integral formulas of Bessel functions
\begin{eqnarray}\label{grOPE3}
\int_0^{2\pi} d\theta  e^{-2ir \cos \theta}&=& 2\pi J_0(2r)\ , \quad
\int_0^{2\pi} d\theta e^{i\theta} e^{-2ir \cos \theta}= - 2\pi i
 J_1(2r)\ ,  \\
\int_0^\infty dr r ^a J_0(2r)&=&  \frac{1}{2}{ \Gamma\left( {1+a\over 2}\right)\over
\Gamma\left({1-a\over 2}\right)}\ , \quad
\int_0^\infty dr r ^a J_1(2r)= {1\over 2} {
\Gamma\left(1+ {a\over 2}\right)\over \Gamma\left(1- {a\over 2}\right)} \ .
\end{eqnarray}

Defining $P^{\mu \nu}$ and $ K^{\mu \nu}$ as in (\ref{defPK}) and after some trivial
manipulations we arrive at the final formula
\begin{eqnarray}\label{bospomcl2}
M^{\mu \nu ;\rho\sigma}&=& {\Gamma\left( \hapr p_1\cdot p_2 -1\right)\over
\Gamma\left(2-\hapr p_1\cdot p_2 \right) } e^{i p \cdot X}
  (Q \cdot \partial X  Q\cdot \bpartial X)^{-\hapr p_1\cdot p_2}\\
& & \left[ (Q\cdot \partial X) P_{\mu \rho} - K_{\mu \rho}
\left(\hapr p_1\cdot p_2-1\right)\right] \times \left[ (Q \cdot \bpartial X)
\bar{P}_{\nu \sigma} - \bar{K}_{\nu \sigma} \left( \hapr p_1\cdot p_2-1\right)
\right]\nonumber
\end{eqnarray}

We show now  how we can derive
(\ref{susyholpomcl2}). We start with (\ref{susyholpomcl}) which, using the standard technique of
rewriting the vertex operators as exponentials in terms of Grassmann variables,
becomes
\begin{eqnarray}\label{susyholpomcl3}
&\he_{1;\mu} \he_{2;\nu} e^{-\varphi(0)}
e^{2i k\cdot X(0)+ 2i
y(\hk_1-\hk_2)\cdot \partial X(0) } (2y)^{2\apr k_1\cdot k_2}
\big(1+y \dot{\phi}(0)\big)&   \nonumber \\
&\Bigg[ -i \hapr \left( {\hk_2^\mu \psi^\nu(0) -\hk_1^\nu
\psi^\mu(0) \over y}\right)+ i \hapr \big(\hk_2^\mu
\dot{\psi}^\nu(0) + \hk_1^\nu \dot{\psi}^\mu(0)\big) + \left(
\partial X^\mu + i\apr  \big(\hk_1 \cdot \psi\big)
\psi^\mu\right) \psi^\nu (0)& \nonumber \\
&- i \hapr {\eta^{\mu\nu} \over y} \left( (\hk_1\cdot \psi) + y
(\hk_1\cdot \dot{\psi})\right)(0) \Bigg] \ . &
\end{eqnarray}
We make now the substitution $Y=y Q \cdot
\partial X$ and the integrations are usual gamma function
integrals. In addition, we use the physical state conditions
$\he_i^\mu \hk_{i;\mu}=0$ to rewrite
\begin{equation} \label{id}
\he_1^\mu \hk_{2;\mu}=\he_1^\mu k_\mu\ , \quad  \he_2^\mu
\hk_{1;\mu}=\he_2^\mu k_\mu \ .
\end{equation}
The final result after the integration over $y$ is
\begin{eqnarray}\label{susyholpomcl4}
\he_{1;\mu} \he_{2;\nu} &&e^{-\varphi(0)} e^{2i k\cdot
X(0)}\Gamma(-1+ 2\apr  k_1\cdot k_2) (Q\cdot
\partial X)^{-1-2\apr k_1\cdot k_2} \times
\nonumber \\ &&\times \Big( \eta^{\mu\nu}(Q\cdot
\partial X)(\hk_1\cdot \psi) +  A^{\mu\nu}_1
+ \eta^{\mu\nu} C + (Q\cdot \partial X)A_2^{\mu\nu} + \dots \Big)
\end{eqnarray}
where
\begin{eqnarray}\label{susyholpomcl5}
&A_1^{\mu \nu}=-2(\hk_1\cdot \psi) \psi^\mu \psi^\nu (-1+2 \apr
k_1\cdot k_2)\ , \quad
A_2^{\mu\nu}=-(\psi^\mu k^\nu-\psi^\nu k^\mu)
\ ,&
\nonumber \\ &\quad
C=(-1 +2 \apr  k_1\cdot k_2) \left(
(\hk_1\cdot \psi) \dot{\phi}+ (\hk_1\cdot \dot{\psi})\right)
\ .&
\end{eqnarray}

\renewcommand{\thesection}{B}
\setcounter{equation}{0}
\renewcommand{\theequation}{B.\arabic{equation}}
\section*{Appendix B: Open-closed mixing pomeron}\label{B}

Assume that the closed string tachyon has only world-volume
momenta, $\apr \hp\cdot D \cdot \hp=\apr \hp\cdot V \cdot \hp=4$.
The relevant OPE takes the form
\begin{eqnarray}\label{mixOPE}
&& : e^{i \hp \cdot X(w)}: \ :e^{i D \hp \cdot X(\bw)}: \ :e^{2i\hk \cdot X(z)}: \; \sim \nonumber \\
&& (w-\bw)^{\hapr \hp \cdot D \cdot \hp} (w-z)^{\apr \hp \cdot
\hk} (\bw -z)^{\alpha'  \hp \cdot D \cdot \hk} :e^{i \hp \cdot X(w) + i D
\hp \cdot X(\bw) + 2i \hk \cdot X(z)}: \ .
 \end{eqnarray}
Using $D\hp =
V\hp =\hp$ and Cartesian coordinates for the upper half-plane $w=x+iy$ as well as
expanding $X(z)$ around
 $x$ and setting $z=x+R$ yields the pomeron operator
\begin{eqnarray}\label{mixOPE2}
&& {\cal V}^{T_c T_o}\sim \int_{-\infty}^{+ \infty} dR \int_{0}^{\infty} dy
y^2 (R^2+y^2)^{\apr  p \cdot \hk} e^{ 2i R \hk \cdot \partial X} \ e^{2
i (\hp +\hk) \cdot X(x)}\ .
\end{eqnarray}
The integration above gives
\begin{equation}\label{mixOPE3}
{\cal V}^{T_c T_o}(\hk, \hk+ V\cdot \hp)\sim (\hk \cdot \partial X)^{-2\apr  \hk
\cdot \hp -4} {\Gamma\left(-{3\over 2} -\apr \hp \cdot \hk\right) \Gamma(
4+2\apr \hp \cdot \hk) \over \Gamma(-\apr\hp\cdot\hk)
\Gamma\left({1\over 2}+ \apr \hp \cdot \hk\right) \Gamma\left({1\over 2}- \apr \hp
\cdot \hk\right)} e^{2 i (V \hp +\hk) \cdot X(x)}\ .
\end{equation}
Finally, we can insert this operator on the disc along with two open
string tachyons to calculate
\begin{equation}\label{mixOPE4}
\left \langle {\cal V}^{T_c T_o}(\hk_3, \hk_3+ V\cdot \hp) \ V^{T_o}(k_1) \ V^{T_o}(k_2)
\right \rangle
\end{equation}
and we get the result (\ref{Regge3t1T}) for $\apr (V\cdot \hp)^2= 4$,
$ 2\apr  \hp \cdot \hk = -s-3$.

\renewcommand{\thesection}{C}
\setcounter{equation}{0}
\renewcommand{\theequation}{C.\arabic{equation}}
\section*{Appendix C: The candidate effective actions}\label{C}

In this appendix we present the two low energy effective actions,
supergravity and DBI, that can serve as a starting point of our
analysis. For the supergravity we consider an expansion $G_{\mu
\nu}=g_{\mu \nu}+ 2\kn h_{\mu \nu} $  in terms of gravitational
fluctuations $h_{\mu \nu}$ about an arbitrary classical background $g_{\mu
\nu}$.   We consider only the NS-NS sector and moreover we ignore
the Kalb--Ramond field for simplicity. We write the action plus
de-Donder gauge fixing terms. Furthermore, we make a field
redefinition to decouple the dilaton from the physical graviton
field, i.e.~we go to the Einstein frame\footnote{We use middle
alphabet Greek letters for bulk indices.} \be h_{\mu \nu} \to
h_{\mu \nu} + g_{\mu \nu} \sqrt{\frac{2}{D-2}} \phi, \ \ \ \ \ \ \
\ \ \ \phi \to \frac{1}{2} g^{\mu \nu} h_{\mu \nu} +
\sqrt{\frac{D-2}{2}} \phi \ee so that the gravity Lagrangian
simply becomes \be \label{Lgrav} L = \sqrt{-g} \left(\frac{1}{4}
g^{\mu \nu} g^{\kappa \rho} g^{\lambda \sigma} \nabla_\mu
h_{\kappa \lambda} \nabla_\nu h_{\rho \sigma} - \frac{1}{2}
h_{\kappa \lambda} h_{\mu \nu} R^{\lambda \mu \kappa \nu} +
\frac{1}{2} g^{\mu \nu}
\partial_\mu \phi
\partial_\nu \phi \right).
\ee
 We will consider only amplitudes involving gravitons since this suffices for our purposes.
Therefore we will drop the (re-defined) dilaton field, since it
decouples from the amplitudes we are interested in.

The abelian DBI action for a D$p$-brane in the Einstein frame ($d=10$) takes the
form
\begin{equation}\label{biact2}
S_{DBI}=-T_p \int d^{p+1}\sigma \  e^{{p-3\over4}\Phi}
\sqrt{-\det(\tG_{\a \b}+e^{-\Phi/2}\tB_{\a \b}+2\pi \apr
\,e^{-\Phi/2}F_{\a \b})} \ .
 \end{equation} The fields
$\tG_{ab}$ and $\tB_{ab}$ are pull-backs
\begin{equation}\label{pullbacks}
\tG_{\a \b}= G_{\mu \nu}\partial_{\a} X^{\mu}
\partial_{\b} X^{\nu}, \qquad \tB_{\a \b}= B_{\mu \nu} \partial_{\a} X^{\mu}
\partial_{\b} X^{\nu}
\end{equation}
 of the bulk fields $G_{\mu\nu}, B_{\mu \nu}$. The fields $ X^\mu(\sigma), \; \mu=0,1,\ldots, d-1$ 
describe
the embedding of the $p$-dimensional world-volume of the brane, which is parametrized
by $\sigma^{\a}, \; a= 0 ,\dots , p$,
in the ambient spacetime. The
two vector fields $ \partial_{\a} X^{\mu}$ and $ \xi_{i}^{\mu}, \;
i=p+1,\dots,9 $ define tangent and normal bundle frames
respectively and satisfy the relations
\begin{eqnarray} \label{DefSub}
\xi_{i}^{\mu}\xi_{j}^{\nu} G_{\mu \nu} = \delta_{i j}\ , \quad
\quad \xi_{i}^{\mu}\partial_{\a} X^{\nu} G_{\mu \nu}=0
\end{eqnarray}
where $ \delta_{ij} $ is the normal bundle metric. Using these
frames we pull-back tensors from  the ambient space to the tangent
and normal bundle.

In the non--abelian action the various brane fields become
matrix-valued functions and the symmetrized trace prescription is applied on the
usual DBI. Moreover, there are extra terms involving commutators
of the fields. We will not give the form of the non--abelian
Lagrangian here since it will not be necessary for what follows.

We choose to work in the static gauge where the world-volume
coordinates  $\sigma^\alpha$ coincide with the bulk coordinates $X^{\mu} $ for $
\mu = 0, \dots, p,  $ which implies
\begin{equation}\label{statgauge}
\partial_\a X^{\b}= \delta_{\a}^{\b}\ , \quad
G_{\mu \nu}(X^\kappa)=G_{\mu \nu}\left(\sigma,X^i(\sigma)\right)\ .
\end{equation}
The fields $X^i $
describe transverse fluctuations of the brane. In this gauge
(\ref{pullbacks}) takes the form
\begin{eqnarray} \label{pullbacks2}
\tilde{G}_{\a \b}= G_{\a \b} + 2 G_{i(\a}
\partial_{\b)} X^i + G_{ij}\partial_{\a}X^i
\partial_{\b}X^j
\end{eqnarray}
and similarly for the antisymmetric tensor field.

Now, we expand the DBI action to second order in the hard fields
$h_{\mu \nu}$, $\lambda_i$ and $f_{\a \b}$. Since we will consider
for simplicity only graviton amplitudes we can set $B_{\mu \nu}=0$
and $\Phi=0$. The various fields are expanded around the soft
background $g_{\mu\nu}, x^i$ and $A_\alpha$
as follows
\begin{eqnarray}\label{fieldexp}
 G_{\mu\nu}=g_{\mu \nu} + 2\kn h_{\mu \nu}\ , \quad
X^i= x^i + {1\over \sqrt T_p} \lambda^i\ , \quad
 \tilde{A}_\a= A_\a + {1\over \sqrt{T_p} 2\pi \apr} a_\a \ .
\end{eqnarray}
Notice that $g_{\mu\nu}$ and $h_{\mu \nu}$ depend on
$X^i$ and therefore the total fluctuation in $G_{\mu \nu}$
includes also $\lambda$.

It will be useful to write down the coupling constants  that
appear in the action in terms of Newton's constant $\kn$, the
coupling of the gauge fields on the brane $g_{Dp}$ and $\apr$.
Moreover we give the relations among the open $g_o$ and closed
$g_c$ string couplings which appear in the definitions of the
string vertex operators. We have the following relations
\cite{Lust:2008qc, Polchinski:1998rr} which will be useful in our
discussion
\begin{eqnarray}\label{constants}
g_c= 2\pi \kn\sim g_s (\apr)^2\ , \qquad && {g_o^2 \over g_c} \sim
(\apr)^2\ , \nonumber \\
 T_{p}\sim {1\over (\apr)^2 g^2_{Dp}}\ , \qquad && g^2_{Dp} \sim g_s (\apr)^{(p-3)/2}\ ,
\end{eqnarray}
where $g_s$ is the asymptotic closed string coupling and we have
ignored the exact numerical coefficients.

For simplicity we will restrict ourselves to the case where the gauge
field are turned off. In other words we
will not consider amplitudes which involve gauge fields either
soft or hard\footnote{Actually, if we keep the gauge fields the formulas
below are written in terns of the open string metric
\cite{Seiberg:1999vs} instead of the induced one.}. The quadratic
action is of the schematic form
\begin{equation}\label{quadDBI}
S_{DBI}^{quad}= S^{gr}_{DBI} + S^{sc}_{DBI}+ S^{mix}_{DBI}\ ,
\end{equation}
where the various contributions take their name from the hard fields
they involve. From the expansion of the square root
\begin{eqnarray}
\sqrt{\det(\tg_{\a \b}+\tilde{h}_{\a \b})}&=& \sqrt{\det \tg_{\a
\b}} \left(1+{1\over2} \tilde{h}_{ \a}^{\a}-{1\over4}\tilde{h}_{ \a \b}
\tilde{h}^{ \b \a}+{1\over8}(\tilde{h}_{ \a}^{\a})^2 +\dots\right)
\end{eqnarray}
we obtain the two graviton action
\begin{equation}\label{qgrav}
S^{gr}_{DBI} \sim {1\over (\apr)^2 g^2_{Dp}} \int d^{p+1}\sigma \
\left( \kn \tilde{h}_{\a \b} T_{DBI}^{\a \b} + \kn^2 \tilde{h}_{\a \b}
\tilde{h}_{\gamma \delta}Z_{DBI}^{\a \b; \gamma \delta}\right)\ ,
\end{equation}
where of course all indices are raised and lowered
with the $\tg_{\a \b}$ metric\footnote{Notice that
$\tg_{\a \b}$ and ${\tilde h}_{\a \b}$ are the pull-backs of $g_{\mu \nu}$ and $h_{\mu \nu}$
evaluated at $\lambda^i=0$.}.
The linear coupling is proportional
to the brane stress energy tensor $T^{\alpha \beta}_{DBI}$. The quadratic
term has a coefficient $Z_{DBI}^{\a \b; \gamma \delta}(\tg)$ with
no symmetry in its indices.

The two scalar action takes the form
\begin{equation}\label{qscalar}
S^{sc}_{DBI} \sim \int d^{p+1}\sigma \ \sqrt{\det \tg} \tg^{\a \b}
g^{ij} \partial_{\a} \lambda_i \partial_\b \lambda_j\ .
\end{equation}
Then, using vielbeins to expand the metric in the normal and
tangent bundle
\begin{equation}\label{vielnorm}
g_{\mu \nu}= e^{\tilde i}_\mu e^{\tilde j}_\nu \delta_{\tilde i \tilde j} + e^\a_\mu e^\b_\nu
\eta_{\a \b}\ , \quad e^{\tilde i}_\mu e^{\a\mu}=0
\end{equation}
and making the field redefinition $\lambda_i = e^{\tilde i}_i
\lambda_{\tilde i}$ we arrive at the simple geometrical action
\begin{equation}\label{qscalar2}
S^{sc}_{DBI} \sim \int d^{p+1}\sigma \ \sqrt{\det \tg}\tg^{\a \b}
D_\a \lambda^{\tilde i} D_\b \lambda_{\tilde i} \ .
\end{equation}
The covariant derivative above is defined using the normal bundle
connection $\omega_{N;\a}^{[ \tilde i \tilde j]}$ (see for instance
\cite{Fotopoulos:2001pt})
\begin{equation}\label{normcovder}
D_\a \lambda_{\tilde i}= \partial_\a \lambda_{\tilde i} + (\omega_{N;\a}^{[\tilde i \tilde j]} -
\partial_\a x^\mu \Gamma^\rho _{\mu \kappa} e^\kappa_{\tilde i} e^{\tilde j}_\rho)
\lambda_{\tilde j}
\end{equation} as it is the natural object which commutes with
the vielbeins $e^{\tilde i}_i$. The action mixing bulk and brane
modes takes the form
\begin{equation}\label{qmix}
S^{mix}_{DBI} \sim {1\over \apr g_{Dp}} \kn  \int d^{p+1}\sigma \
\sqrt{\det \tg} \tg^{\a \b} g^{\mu \nu} \left( \partial_\a
\lambda_\mu
\partial_\b x^\kappa h_{\kappa \nu} + \lambda_\mu
\partial _{\a} x^\kappa \partial _{\b} x^\lambda \partial_\nu h_{\kappa \lambda} \right)\ .
\end{equation}
This  can also be written in an equivalent form using the normal bundle
vielbeins. The second term in the action above originates from
Taylor expanding the metric in (\ref{statgauge}) around the
background in (\ref{fieldexp})
\begin{equation}\label{Taylorg}
G_{\mu \nu} (\sigma, X^i) = G_{\mu \nu}(\sigma, x^i) + \lambda^i
\partial_i G_{\mu \nu} (\sigma, x^i) +\dots \ .
\end{equation}  The full action needed for
our analysis is given by the sum of the actions in (\ref{Lgrav}) and 
(\ref{quadDBI}).

    \end{document}